\documentclass[twocolumn,floatfix]{aastex63}
\usepackage{algorithm}
\usepackage{algpseudocode}
\usepackage{verbatim}
\usepackage{hyperref}
\usepackage{amsmath}
\usepackage{tabularray}
\usepackage{changepage}
\usepackage{pifont}
\newcommand{\cmark}{\ding{51}}%
\newcommand{\xmark}{\ding{55}}%
\usepackage{xspace} 
\xspaceaddexceptions{]\}}

\usepackage{xcolor,colortbl}

\definecolor{Gray}{gray}{0.85}
\definecolor{applegreen}{rgb}{0.55, 1, 0.0}
\definecolor{alizarin}{rgb}{0.82, 0.1, 0.26}
\definecolor{lightred}{rgb}{1, 0.4, 0.4}
\definecolor{lightgreen}{rgb}{0.56, 0.93, 0.56}
\definecolor{lightyellow}{rgb}{0.78, 0.665, 0.48}
\definecolor{LightCyan}{rgb}{0.88,1,1}
\newcolumntype{a}{>{\columncolor{Gray}}c}
\newcolumntype{b}{>{\columncolor{white}}c}
\newcolumntype{g}{>{\columncolor{applegreen}}c}
\newcolumntype{r}{>{\columncolor{alizarin}}c}
\newcommand{\spender}{\texttt{spender}}
\newcommand{\cigale}{\texttt{CIGALE}}
\newcommand{\CIGALE}{\texttt{CIGALE}}
\newcommand{\prospector}{\texttt{prospector}}

\newcommand{\fsps}{\texttt{fsps}}

\usepackage{braket}

\shorttitle{The optical and infrared are connected}
\shortauthors{C.~K.~Jespersen et al. }

\begin{document}

\title{The Optical and Infrared Are Connected}

\correspondingauthor{Christian Kragh Jespersen}
\email{ckragh@princeton.edu}

\author[0000-0002-8896-6496]{Christian Kragh Jespersen}
\affiliation{Department of Astrophysical Sciences, Princeton University, Princeton, NJ 08544, USA}

\author[0000-0002-8873-5065]{Peter Melchior}
\affiliation{Department of Astrophysical Sciences, Princeton University, Princeton, NJ 08544, USA}
\affiliation{Center for Statistics and Machine Learning, Princeton University, Princeton, NJ 08544, USA}

\author[0000-0002-5151-0006]{David N. Spergel}
\affiliation{Department of Astrophysical Sciences, Princeton University, Princeton, NJ 08544, USA}
\affiliation{Center for Computational Astrophysics, Flatiron Institute, 162 5th Avenue, New York, NY 10010, USA}

\author[0000-0003-4700-663X]{Andy D. Goulding}
\affiliation{Department of Astrophysical Sciences, Princeton University, Princeton, NJ 08544, USA}

\author[0000-0003-1197-0902]{ChangHoon Hahn}
\affiliation{Department of Astrophysical Sciences, Princeton University, Princeton, NJ 08544, USA}

\author[0000-0001-9298-3523]{Kartheik G. Iyer}
\affiliation{Columbia University, Columbia Astrophysics Lab, 550 W 120th St, New York, NY 10010, USA}

\begin{abstract}
Galaxies are often modelled as composites of separable components with distinct spectral signatures, implying that different wavelength ranges are only weakly correlated. They are not. We present a data-driven model which exploits subtle correlations between physical processes to accurately predict infrared (IR) WISE photometry from a neural summary of optical SDSS spectra. The model achieves accuracies of $\chi^2_N \approx 1$ for all photometric bands in WISE, as well as good colors. We are able to tightly constrain typically IR-derived properties, e.g., the bolometric luminosities of AGN and dust parameters such as $\mathrm{q_{PAH}}$. We also test whether current SED-fitting methods reproduce such panchromatic relations, but find their predictions biased and overconfident, likely due to model misspecification, with correlated biases in star-formation rates and AGN luminosities being most evident.  To help improve SED models, we determine which features of the optical spectrum are responsible for our improved predictions, and identify several lines (CaII, SrII, FeI, [OII] and H$\alpha$), which point to the complex chronology of star formation and chemical enrichment being incorrectly modelled.
\newline
\end{abstract}
\section{Introduction}
\label{sec:intro}

Understanding the properties of galaxies and their evolution is one of the major focus point for the upcoming decades of astronomy \citep{Astro2020_decadal}, as galaxies are both fascinating in their own right, and a fundamental tool for cosmology \citep{ Hahn2023_simbig, DESI_BAO}.
In the current paradigm of galaxy formation theory, galaxies develop through hierarchical growth, eventually forming many distinct components, such as the disk, bulge and central super-massive black hole (SMBH) \citep{SomervilleDave2015_review}. These components are often modelled as evolving independently from each other, e.g. the stars in the outer disk and the circumgalactic medium are not directly affecting the SMBH. This assumption is often made implicitly, as modelling the interaction of different galaxy components is beyond our current understanding. Large-scale galaxy simulations are getting ever closer to being able to explore the interplay of these components, but for now they still rely on insufficient, resolution-limited subgrid models \citep{Wright2024_baryoncycle}. The implicit assumption of \textit{separability} of many of the components of a given galaxy also implies that different regions in wavelength space are largely separable, since the different components emit across different wavelengths.

The assumption that different components are approximately separable is propagated through to galaxy spectral energy distribution (SED) fitting codes, the most common method of deriving the physical properties of galaxies \citep{Noll2009_cigale, ConroyGunn2010_SSPII_comparing_observations_to_models, Conroy2013_SED_Fitting_review, Johnson2021_prospector, Pacifici2023_SED_fitting, Iyer2025_review}. SED models comprise templates for separate components, which usually do not interact, except through a few constraints, e.g. to balance the amount of energy absorbed by dust at short wavelengths with its re-emission at longer wavelengths \citep{daCunha2008_magphys}. This constraint is based on the success of simple relations such as the star-forming $\mathrm{H}\alpha$--FIR connection \citep{Inoue2002_SFR_dust, Kewley2002_Halpha_IR_SFR}. Early works, such as \cite{Rieke1978_SeyfertIR, RiekeLebofsky1979_IR_review}, also discussed the possible connections between optical properties and IR emission. As the astronomical community is now in possession of many hundreds of thousands of objects with coverage from the ultraviolet (UV) to the mid-infrared (MIR) \citep[e.g., GSWLC, COSMOS2020,][]{Salim2016_GALEX_SDSS_WISE, Weaver2022_COSMOS}, as well as tools for large-scale data analysis, it is possible to find more general relations \emph{if these relations exist}. 

In this paper we explore correlations across broad wavelength ranges. We do this by connecting the optical spectroscopy from the Sloan Digital Sky Survey \citep[SDSS;][]{York2000_SDSS, Gunn2006_SDSS} to near- and mid-infrared (NIR/MIR) photometry from the Wide-field Infrared Survey Explorer \citep[WISE;][]{Wright2010_WISE_overview} satellite. SDSS conducted spectroscopy of half a million galaxies between 3800 and 9200 Å at R $\sim 2000$ resolution, while WISE conducted all-sky four-band photometry from 3 to 27 $\mathrm{\mu m}$. SDSS was designed to be sensitive to the main stellar population in a typical galaxy, whereas WISE covers the transition from emission being dominated by older stellar populations to dust tori from AGN and asymptotic giant branch (AGB) stars, polycyclic aromatic hydrocarbons (PAHs) and hot dust. The combination of these two datasets is therefore ideal to investigate whether galaxy components are truly separable. We compare our results to results from two state-of-the-art SED fitting codes, \cigale~\citep{Boquien2019_CIGALE, Yang2020_CIGALE_xray} and \prospector~\citep{Leja2017_prospector_use_1, Johnson2021_prospector}.

Testing a physical model's ability to predict quantities not directly included in the fit is a well-established validation approach. For example, \citep{Leja2017_prospector_use_1} validated the \prospector-$\alpha$ model by comparing photometry-only predictions against spectroscopic quantities excluded from their fits, including H$\alpha$ luminosities and Balmer decrements. Our approach is analogous: we ask whether models constrained by optical data produce physically reasonable predictions in the infrared, a natural self-consistency test for any model providing a panchromatic representation of galaxy emission.

The paper is structured as follows. In \S \ref{sec:data} we describe the SDSS and WISE data used in this study, as well as the preprocessing of the data. In \S \ref{sec:methods} we introduce the methods used to create the empirical mapping from optical spectra to IR photometry, as well as our setup for using \cigale~and \prospector. \S \ref{sec:results} shows results from our empirical mapping and the SED fitting codes. We furthermore demonstrate that physical quantities typically derived from the IR can be robustly predicted using the optical spectrum alone, directly showing that the current picture of separable components is incorrect. In \S \ref{sec:discussion}, we investigate what information is used for these accurate predictions and where SED models must improve. We furthermore demonstrate that the separability assumption in SED models leads to biases and imprecision in parameter inference. In \S \ref{sec:conclusion} we conclude on our work and outline future directions.

\section{Data}
\label{sec:data}

The two data sources, SDSS and WISE, cover wavelengths spanning a factor of $\approx60$ ($0.4-24 \mu$m). The total dataset consists of $5.1\cdot10^5$ crossmatched objects at $z \leq 0.5$. 

\subsection{SDSS spectra}
\label{subsec:sdss_data}

\begin{figure}
  \centering
  \includegraphics[trim={0.2cm 0.0cm 0cm 0.cm},clip,width=0.95\linewidth]{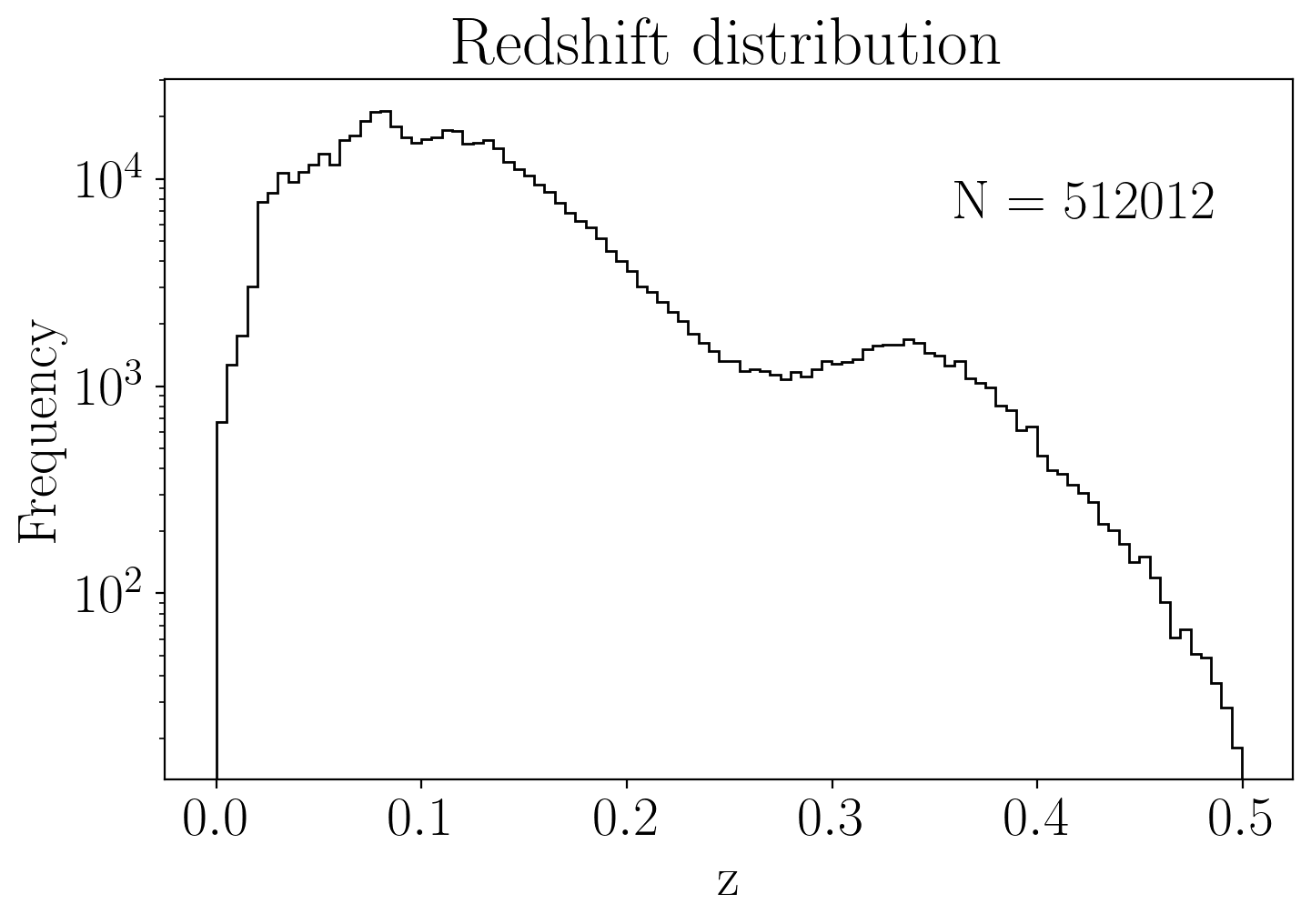}
  \caption{The redshift distribution of the objects used in this work.}
  \label{fig:redshift_demonstration}
\end{figure}
We utilize $5.1\cdot10^5$ spectra from the Main Galaxy Sample of the Sloan Digital Sky Survey \citep[SDSS-II]{Strauss2002_sdss_spec_selection}. The spectroscopic part of the main galaxy sample from the SDSS survey was a magnitude-limited survey with high completeness. The targets were relatively bright galaxies with $m_{AB,i} \leq 17.8$ \citep{ Strauss2002_sdss_spec_selection}. Figure \ref{fig:redshift_demonstration} shows the distribution of redshifts.
\begin{table}
\caption{Classifications of the spectra utilised in this work. ``None'' refers to any spectrum not falling into any other category, which mainly covers quiescent galaxies.}
\label{tab:sdss_classifications}
\centering
 \begin{tabular}{| c | c | c | c | c|} 
 \hline
  & None & Star-forming & Starburst & AGN \\ 
 \hline
  Narrow line &  318,795 & 148,750 & 23,754 & 10,731 \\ 
 \hline
  Broad line & 7,922 & 673 & 37 & 1,350 \\ 
 \hline
 \end{tabular}
\end{table}

Due to the magnitude-limited nature of the survey, the Main Galaxy Sample includes a high diversity of galaxies. The spectroscopic classifications (as determined by SDSS) are outlined in Table \ref{tab:sdss_classifications}.

The high brightness limit allowed SDSS to generally obtain very accurate fluxes, since the main continuum foreground, the airglow emission of the night sky, varies between 20th and 22nd AB magnitude in the i-band \citep{RoachGordon1973, Jespersen2024_SuNSS_SPIE}. There are naturally also many stronger emission lines in the airglow, reaching up to 12th AB magnitude \citep{Jespersen2024_SuNSS_SPIE}, but these skylines are masked in our analysis. The SDSS spectra were taken in fixed aperture fibers, with a diameter of $3''$, different from the profile-fit WISE photometry. The implications of this difference will be evaluated and discussed in \S \ref{subsec:results_breakdown_size}. 

SDSS also erroneously targeted stars and other outlier targets, such as chance alignments of multiple galaxies. These are removed first based on the SDSS spectroscopic classifications, and second based on the outlier analysis of \cite{Liang2023_spender_outliers}.

For the data-driven model, we first preprocess the spectra using the \spender~spectral autoencoder \citep{Melchior2023_spender, Liang2023_spender}. \spender~is publicly available and can be \texttt{pip}-installed.\footnote{See \href{https://github.com/astrockragh/IR_optical_demo}{https://github.com/astrockragh/IR\_optical\_demo} for a demonstration.} \spender~encodes each SDSS spectrum into a six-dimensional latent space, from which the spectrum can be reconstructed. This autoencoding step makes the inference more robust and significantly faster. An expanded description of \spender~can be found in Appendix \ref{appsec:autoencoder}. A discussion of the implications of using autoencoded spectra as the basis for analysis can be found in \S \ref{subsec:science_with_latents}.

\subsection{WISE photometry}
\label{subsec:wise_data}
The Wide-field Infrared Survey Explorer \citep[WISE][]{Wright2010_WISE_overview} mapped the sky at 3.4, 4.6, 12, and 22 $\mu$m with angular pixel resolutions of 6.1'', 6.4'', 6.5'' and 12.0'', taking up to 100 exposures of each region of the sky. However, this was not uniformly done, leading to very different detection limits for different regions of the sky. The single-exposure $5\sigma$ point source sensitivities were 0.08, 0.11, 1 and 6 mJy in unconfused regions in the four bands, corresponding to Vega magnitudes of 16.8, 15.6, 11.3, and 8.0.\footnote{ \href{https://wise2.ipac.caltech.edu/docs/release/allsky/}{https://wise2.ipac.caltech.edu/docs/release/allsky/} }
We follow the literature in using the original Vega magnitudes when dealing with WISE data, as well as using W1, W2, W3 and W4 to refer to the WISE bands in order of increasing wavelength. We use the zero-point flux densities identified by \cite{Wright2010_WISE_overview} to convert from magnitudes to flux density in Jansky.\footnote{See \href{https://wise2.ipac.caltech.edu/docs/release/allsky/expsup/sec4\_4h.html}{\nolinkurl{https://wise2.ipac.caltech.edu/docs/release/allsky/expsup/sec4\_4h.html}} } 

\begin{figure*}
  \centering
  \includegraphics[trim={0.25cm 0.0cm -1cm 0.cm},clip,width=0.95\linewidth]{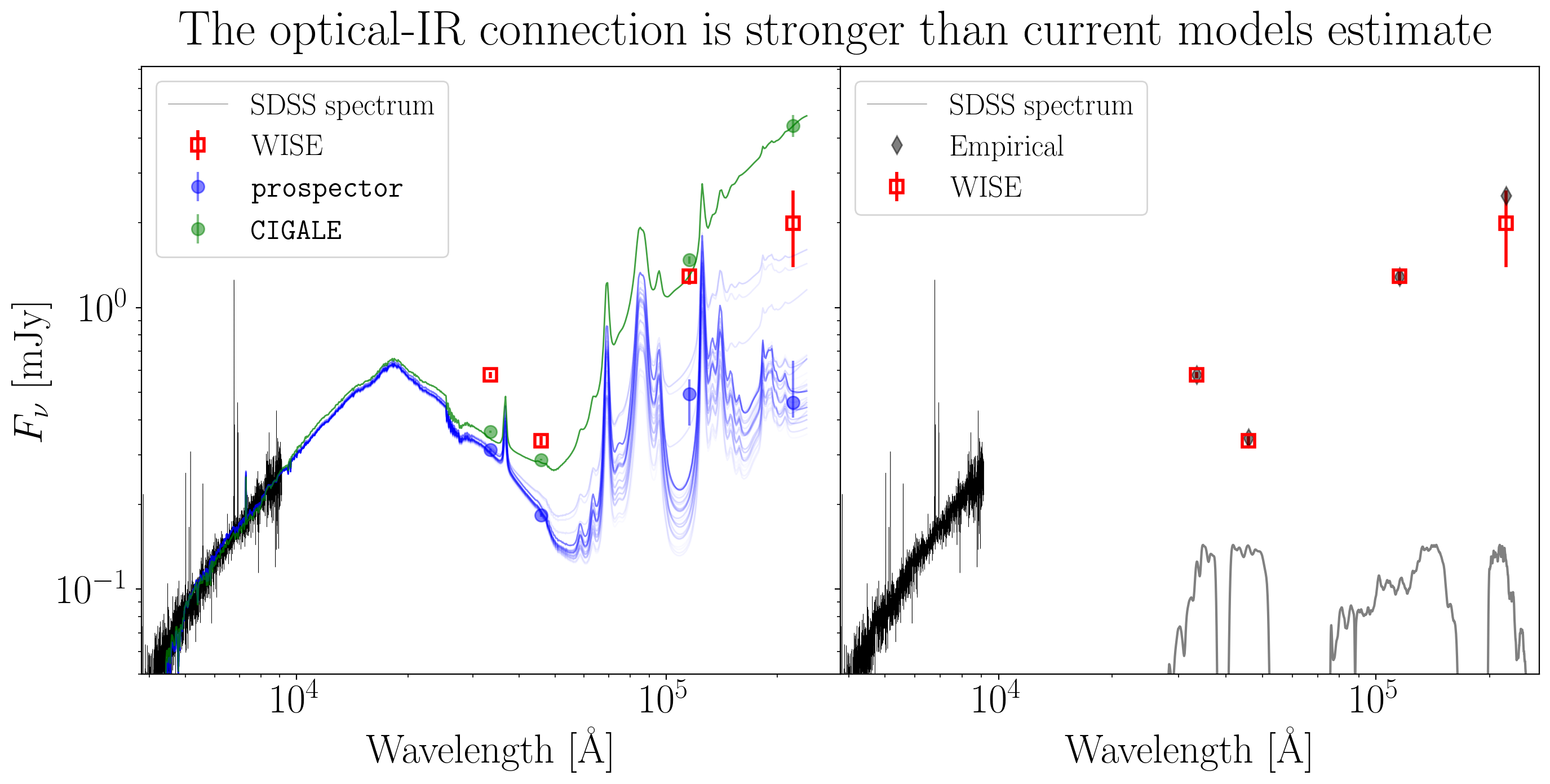}
  \caption{A demonstration of predicting WISE photometry (red markers) from the SDSS spectrum (black), shown in observed wavelength. Although they fit the optical spectrum perfectly, the predictions from the SED models (left panel) have typical deviations of $\sim 10 \sigma$, whereas our empirical approach (right panel) produces typical deviations of $\sim 1 \sigma$. Crucially, the SED models are very confident in their biased extrapolation. That the photometry can empirically be predicted to $\sim 1 \sigma$ implies that IR photometry is fully determined by the optical spectrum. The multiple lines for \prospector~are the results of 100 draws from the posterior. The model errors of \cigale~are usually smaller than the points themselves.}
  \label{fig:demonstration}
\end{figure*}

The WISE catalogues were downloaded from the SDSS SkyServer ($\texttt{WISE\_allsky}$ table), along with the crossmatched SDSS-WISE ID table ($\texttt{WISE\_xmatch}$ table).\footnote{Downloaded on 2024-02-26 from \href{https://skyserver.sdss.org/dr16/en/help/browser/browser.aspx?cmd=description+WISE\_allsky+U\#&&history=description+WISE\_allsky+UURL}{SDSS SkyServer webpage} } The crossmatching is done astrometrically with a maximum distance of 2'' and requires a detection in both W1 and W2. The crossmatching leaves out a small fraction of SDSS spectra, but the vast majority is successfully detected by both SDSS and W1/W2. It is discussed further in Appendix \ref{appsec:additional_data}. An example of crossmatched data, along with predictions from the models described later, is shown in Figure \ref{fig:demonstration}.

\section{Methods}
\label{sec:methods}

\subsection{Empirical mapping}
\label{subsec:mlp}

For the empirical mapping we use a simple Multi-Layer Perceptron (MLP). This fundamental neural network building block consists of sequential linear matrix transformations, interspersed with non-linear activation functions. The activation functions have a simple analytic function, the Rectified Linear Unit \citep[ReLU;][]{agarap2019_ReLU}, defined as $\mathrm{ReLU{(x)} = max(0,x)}$.

We make use of the compressed \spender~latent variables\footnote{The latents are a complex, non-linear combination of the input features. For an analysis of the most important features driving the latents, see \cite{Melchior2023_spender}.} from the autoencoded spectra and not the spectra themselves, so we only have eight input parameters (six \spender~latents, a normalization constant, and the redshift). The normalization is a scalar, defined as the median flux in the 5300-5850Å range, used to increase stability in the \spender~encodings. We apply the normalization constant after the output layer, meaning that this approach theoretically can work for arbitrarily faint or bright galaxies. The network used is extremely small for typical machine learning methods, having only three layers and 20 hidden states, for a total of 684 parameters. 

Despite the MLP being our choice for constructing an empirical mapping, it is important to highlight that many other methods can be used with almost equivalent accuracy, such as high-dimensional splines, or any decision tree method. Even using a simple linear regression starting from the \spender~latent variables renders performance only about 40$\%$ worse, as measured by the $\chi^2$.

To optimize our mapping, we minimize a Gaussian Negative Log-Likelihood loss function, defined as:

$$ \mathcal{L}(\mathbf{y}, \hat{\mathbf{y}}, \boldsymbol{\sigma}) \equiv \frac{\ln (\boldsymbol{\sigma})}{2}+\frac{1}{2} \left(\frac{\mathbf{y}-\hat{\mathbf{y}}}{\boldsymbol{\sigma}}\right)^2$$

\noindent where $\mathbf{y}$ is the target vector (measured WISE photometry), $\hat{\mathbf{y}}$ is the network prediction vector (predicted WISE photometry), and $\boldsymbol{\sigma}$ is the uncertainty vector (measured WISE photometric errors). All quantities are in Vega magnitudes.
This likelihood is not strictly appropriate, since it assumes that errors are Gaussian. However, the Gaussian is most likely a reasonable approximation to the best possible likelihood, and since the WISE errors are reported as Gaussian errors, it is the most natural choice. For objects without detections, we set the uncertainty to be $10^4$ Vega magnitudes, effectively allowing the likelihood to discard these data points.

The empirical model is optimized on a subset of the data, typically known as a \textbf{training set}. Here the training set consists of 80\% of the total data. To prevent effects from overfitting, all tests shown in the rest of the paper are performed on a held-out subset of the data, known as the \textbf{test set}. The test set comprises 20\% of the total data. While validating the model set-up and training procedure, a subset of the training data was used as a \textbf{validation set}. The validation set comprises 10$\%$ of the total data and is merged into the training set after all model parameter choices are made. The training, validation, and testing sets are discussed further in Appendix \ref{appsec:train_test_split}.

\subsection{\cigale}
\label{subsec:cigale_setup}

The Code Investigating GALaxy Emission\footnote{\href{ https://cigale.lam.fr}{ https://cigale.lam.fr}} \citep[\cigale,][]{Noll2009_cigale, Boquien2019_CIGALE} is a Bayesian SED-fitting code designed to estimate the physical properties of galaxies. Here, we provide a brief summary of the tool (see \cite{Boquien2019_CIGALE} for a detailed description). \cigale~models the spectra of the different emitting components separately and combines them with dust attenuation and emission. Each spectral module is independent from every other module, as are the parameters within each module. Here, we use six modules, covering the stellar populations, IMF, nebular emission, dust attenuation, dust emission and AGN contribution. \cigale~creates a pre-computed grid of spectra according to a set of user-specified possible values for each parameter for each module. The modules, parameters, and the allowed parameter values are detailed in Table \ref{tab:cigale_params}. The pre-specified values function as an effective prior for the code. Although \cigale~uses a grid of discrete values for evaluation, the output space is nonetheless continuous since \cigale~performs a pseudo-Bayesian interpolation between these discrete grid values based on the likelihood.

\begin{table*}[t!]
  \centering
  \caption{A list of inputs for \cigale. The selection of the modules is discussed in the main text.}
  \label{tab:cigale_params}
  \begin{tabular}{lcc}
  \hline
  \hline
   \cigale~Module& Parameter Description & Possible Values \\
   \hline
  sfhdelayed & E-folding time of main stellar pop [Myr] & 500, 1000, 2000, 5000 \\
       & Main stellar pop. age [Myr] & 100, 500, 1000, 2500, 5000, 10000 \\
       & Burst mass fraction $f_{\mathrm{burst}}$ & 0.0, 0.1, 0.2 \\
       & E-folding time of burst pop [Myr] & 50, 100, 250, 500 \\
  \hline
  bc03 & IMF & Salpeter, Chabrier \\
   & Metallicity & 0.004, 0.02, 0.05 \\
    & Time [Myr] between young and old stellar populations & 100, 300, 1000, 3000 \\
  \hline
  nebular & log(U) (ionizing field strength) & -4.0, -2.0, -1.0 \\
    & Electron density [$\mathrm{cm}^{-3}$] & 10, 100, 1000 \\ 
    & Gas metallicity & 0.004, 0.02, 0.051 \\
    & $f_{\mathrm{esc}}$ (Lyman cont. escape fraction) & 0.0, 0.5 \\
    & $f_{\mathrm{abs}}$ (Lyman cont. absorption fraction) & 0.0, 0.5 \\
    & Line width [km/s] & 200, 400 \\
  \hline
  dustatt\_modified\_CF00 & $\mathrm{A_{V,ISM}}$ & 0.1, 0.35, 0.65, 1.0\\
   
   & $\alpha_{\mathrm{ISM}}$ (Power-law index of ISM attenuation) & -0.7 \\
   & $\alpha_{\mathrm{BirthCloud}}$ (Power-law index of birthcloud attenuation) & -1.3 \\
   \hline
  dl2014 & $\mathrm{q_{PAH}}$ (PAH mass fraction) & 1.12, 2.50, 3.90, 5.26, 7.32.\\
   & $\mathrm{U_{min}}$ (minimum UV intensity) & 0.25, 1.0, 4.0, 17, 40 \\
   & $\alpha$, $\left(dU / dM \propto U^\alpha\right)$ & 1.0, 2.0, 3.0 \\
   & $\gamma$ (Frac. illuminated from $\mathrm{U}_{\min }$ to $\mathrm{U}_{\max })$ & 0.1, 0.5, 0.9 \\
  \hline
  skirtor &$\tau_{\mathrm{9.7 \mu m}}$ (optical depth at 9.7 $\mu$m) & 7, 9, 11 \\
      & $p_l$ (radial dust density power law index) & 0.0, 1.0 \\
      & q (polar dust density gradient) & 0.0, 1.0\\
      & $\theta$ (opening angle) & 10, 30, 50, 80\\
      & $R_{\mathrm{out}}/R_{\mathrm{in}}$ (outer/inner torus radius ratio) & 10, 20, 30 \\
      & i (inclination/viewing angle) & 0, 20, 40, 60, 80\\
      & fracAGN & 0.0, 0.01, 0.1, 0.2, 0.4, 0.8, 0.9, 0.95 \\
      & polar E(B-V) [mag] & 0.03, 0.3, 1.0, 1.5\\
  \hline
  \hline
  
  \end{tabular}
\end{table*}

\subsubsection{Synthetic photometry}

Since \CIGALE~does not have a direct spectrum-fitting mode, it is necessary to convert our spectra into synthetic photometry. To this end, we define box-car filters across the SDSS wavelength range and integrate over the filters to get the synthetic photometry in a given filter. For the uncertainties on the synthetic photometry, we follow the procedure of \cite{Casagrande2014_how_to_make_synthetic_photometry} and \cite{Gaia2023_synthetic_photometry}. However, the exact method of calculating the uncertainties is unimportant, since \cigale~recommends adding 10$\%$ of the flux in quadrature to the error, which dominates the error budget.\footnote{If this ``softening'' is turned off, the fitting often fails, since the likelihood landscape is too sharply peaked.}

We test different numbers of filters ($\{5, 10, 20, 50, 100\}$) and find that 10 or 20 filters usually perform best, with 5 providing weaker constraints, and 50 and 100 often resulting in \cigale~being unable to fit the spectrum. Between 10 and 20 filters there is not much difference, and we therefore use 10 for the sake of simplicity and computational speed.

Emission lines require a bit more consideration. We test explicitly including line strength measurements, focusing on H$\alpha$, H$\beta$, [O III]5008, [N II]6550 and [N II]6585, but it makes very little difference. Since \cigale~does not handle high equivalent width (EW) lines well, we mask out lines which lie above the 99th percentile of EW in any of the aforementioned lines. However, this makes no detectable difference.

\subsubsection{\CIGALE~modules}
\cigale~requires pre-defining a set of modules to use (see Table \ref{tab:cigale_params}). Since the expressivity of \cigale~is entirely controlled by these modules, it is important to choose the correct set \citep{Bellstedt2024_SED_choices}. 
We choose which modules to use by evaluating their ability to \textit{fit} both the optical SDSS spectra and the WISE photometry. 
For parametrizing our star-formation history (SFH), we test the double exponential (\texttt{sfh2exp}), the delayed (\texttt{sfhdelayed}) and periodic (\texttt{sfhperiodic}) SFH modules, finding the first two to be vastly better than the third, with a slight preference for the delayed SFH.

For parametrizing the IMF, we test the models of \citet[][BC03]{Bruzual2003_bc03} and \citet[][M2005]{Maraston2005_m2005_ssp}, but find no significant difference in their ability to provide templates that can fit the IR photometry, despite the M2005 templates paying specific attention to thermally pulsating AGB stars. We therefore use the BC03 module, since it is the default used by \cigale. There is only a single module for parametrizing nebular emission.

For the dust law, we use the model proposed by \citet[][CF00]{CharlotFall2000_dustlaw} and modified by \cite{LoFaro2017_modified_dust_law}. CF00 provides a more detailed model for how the dust around old and young stellar populations behave. We also test the \cite{Calzetti2000} dust law, but find that the modified CF00 dust law provides better fits. 

The two most important modules for characterizing the IR emission of a galaxy are the ones controlling dust emission and AGN activity \citep{Treyer2010_MIR}. The model by \cite{Dale2014} attempts to unify these two in a single, two-parameter model, but it is unfortunately a significantly weaker model than the combination of the PAH/dust emission model by \cite{Draine_Li_2007_PAH} and the SKIRTOR AGN model\footnote{The model can be explored at \href{https://skirtor.streamlit.app/}{https://skirtor.streamlit.app/}} by \cite{Stalevski2016_skirtor}, which together can almost always provide fits to the WISE photometry with $\chi^2_N\approx 1$. 

\subsection{\prospector}
\label{subsec:prospector_setup}

\prospector~is a modern and widely used SED fitting code which implements a more directly Bayesian approach than that taken by \cigale. Instead of a grid of possible values, \prospector~has a full prior and posterior distributions. \prospector~fits using either MCMC \citep{DFM2013_emcee} or nested sampling \citep{Speagle2020_dynesty}. \prospector~also distinguishes itself by its full spectrum-fitting mode, making it the ideal choice for investigating the behaviour of SED models when presented with optical spectra, the most informative data type we currently possess \citep{Johnson2021_prospector}. \prospector~has been applied by its development team to model data ranging from the far-UV ($\approx 0.1 \mu m$) to the sub-mm ($\approx 300 \mu m$) \citep{Leja2017_prospector_use_1, Leja2019_beyond_UVJ}. \prospector~is built atop \fsps~\citep{ConroyGunn2009a_SSPI_uncertain_parameters, Conroy2010_fsps, DFM2014_fsps_python}.\footnote{\href{https://dfm.io/python-fsps/current/}{https://dfm.io/python-fsps/current/}} Since \fsps~contains a large amount of possible parameters, a subset of the parameters is usually specified as free, whereas others, believed to be well known or of no importance, are kept fixed. Two models have been published by the \prospector~team, the \prospector-$\alpha$ model \citep{Leja2017_prospector_use_1} and the \prospector-$\beta$ model \citep{Wang2023_prospector_beta}. The $\alpha$ model is optimized for deriving the physical parameters of galaxies and we therefore base our model on the \prospector-$\alpha$ model, described in \cite{Leja2017_prospector_use_1, Leja2019_older_Q}. All modules and priors used are detailed in Table \ref{tab:prospector_params_priors}. 
Although many aspects of the model are similar to \cigale, there are a few important differences. In contrast to \cigale, \prospector~implements non-parametric SFHs, which is thought to improve the quality of the recovered SFRs \citep{CarnallLeja2019_measuring_SFHs_parametric, LejaCarnall2019_measuring_SFHs_NONparametric}. The AGN model used by \prospector~is that of \cite{Nenkova2008_AGNtorusI}.
We want to stress that although \cigale~and \prospector~are different, they also share a large amount of their basic assumptions, which means that any assessment of ``model uncertainty'' cannot simply be done by comparing different models \citep{Pacifici2023_SED_fitting}. The models share similar SED templates (e.g. those of \cite{Draine_Li_2007_PAH} for dust and PAH emission), as well as the lack of constraints on physically connected parameters (e.g. between star formation, metallicity and $\mathrm{q_{PAH}}$). 

\section{Results}
\label{sec:results}

\begin{figure*}
  \centering
  \makebox[\textwidth][c]{ \includegraphics[width=0.8\linewidth]{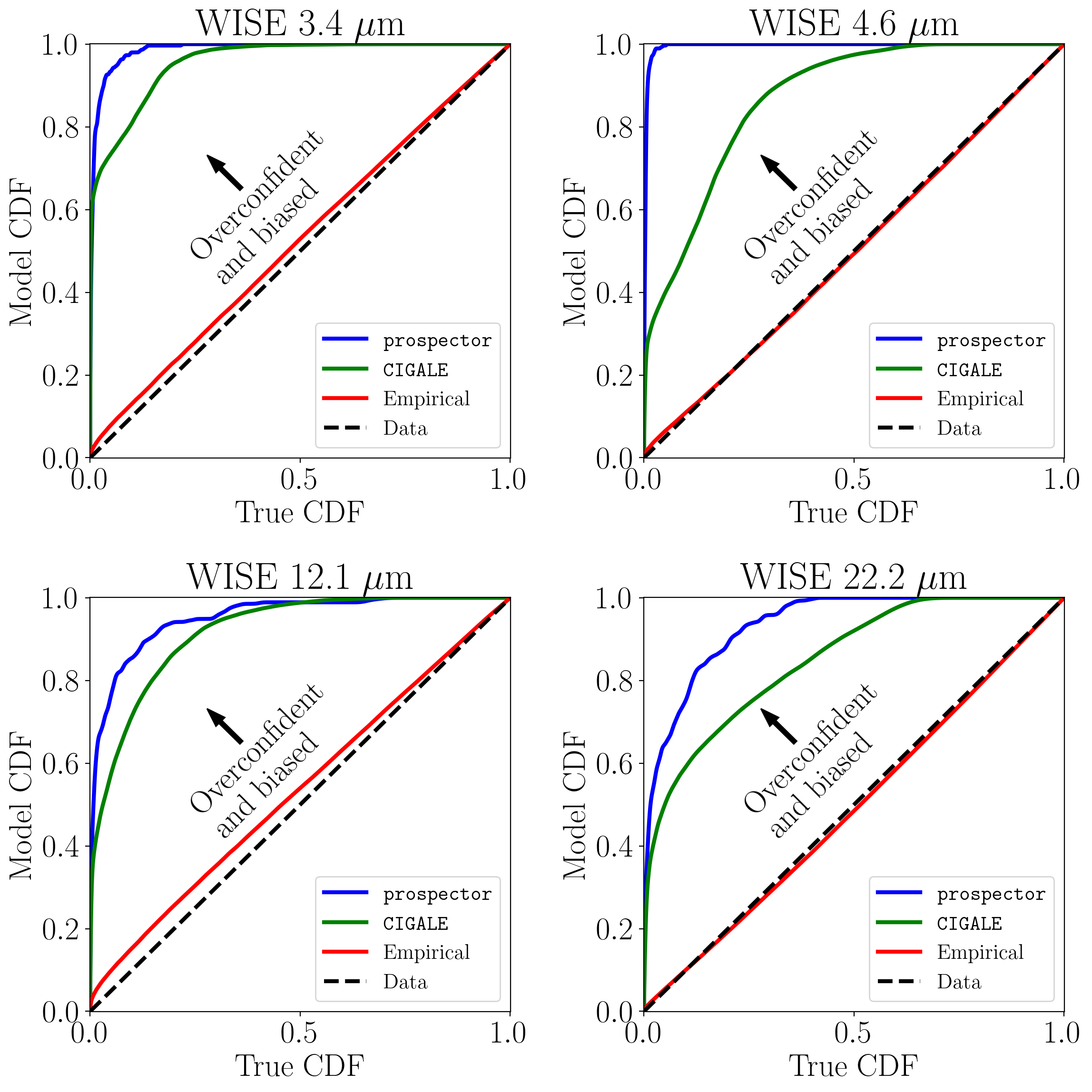} }
  \caption{PP-plot showing the cumulative probability that a sample from the posterior from the empirical model, \prospector, or \cigale~will be consistent with the data. A summary of the mean probabilities can be found in Table \ref{tab:mean_p_fit}. This metric is generally insensitive to catastrophic errors. Additional precision figures are shown in Appendix \ref{appsec:more_figures_chi2}. }
  \label{fig:PP_plot}
\end{figure*}

Here we present results for the task of predicting IR photometry from optical spectra. All results presented will concern only data points with detections in any given band.

To evaluate the performance, we first introduce the reduced $\chi^2$ ($\chi^2_N$), a widely used metric, defined as 

\begin{equation}
  \chi^2_N = \frac{1}{N}\sum_i^N \chi^2,~\chi = \frac{\Delta y_i}{\sigma_i} 
\end{equation}

\noindent where $N$ is the number of data points, $\Delta y_i$ is the residual between the truth and prediction, and $\sigma_i$ is the measurement uncertainty. The $\chi^2_N$ are summarized in Table \ref{tab:mean_chi2_fit}. The $\chi^2_N$ for the \textit{predictions} of the SED-fitting codes are very high, even though both \cigale~and \prospector~\textit{fit} the optical to $\chi^2_N \approx1$. \footnote{See Appendix \S \ref{appsec:optical_fit_quality} for the optical fit quality.} Our empirical model results in $\chi^2_N \approx1$, and we observe no strong dependence on how we subdivide our sample (see Appendix \ref{appsec:sample_split}) or for single extreme galaxies like Arp220 (see Appendix \ref{appsec:arp220}). The only subclass of galaxies with a slightly elevated $\chi^2_N$ is AGN.

The reduced chi-square is a very popular method for model assessment and comparison in astronomy, but it has important caveats that are worth keeping in mind \citep{Andrae2010_chi2_N}. A perfect set of predictions would have $\chi^2_N = 1$ if the underlying $y$ are independent and (unit) normally distributed, neither of which perfectly hold for our data. For example, the $\chi^2_N$ will be inflated if the WISE errors are underestimated \citep{Lang2016_wise_sdss}. However, since WISE errors are reported as Gaussian errors, the reduced chi-square is an appropriate metric.

\begin{table}[h!]
\caption{Mean $\chi^2_N$ of the data under the posterior. Full distributions are shown in the Appendix. This metric is dominated by catastrophic errors. Bolded values highlight the best-performing model for each band.}
\begin{adjustwidth}{-2cm}{}
\label{tab:mean_chi2_fit}
\centering
 \begin{tabular}{| c | c | c | c | } 
 \hline
  & \prospector & \cigale & Empirical \\ 
 \hline
  W1 &  1,441 & 88.1 & \textbf{1.33} \\ 
 \hline
   W2 &  5,936 & 68.2 & \textbf{1.17} \\ 
 \hline
   W3 &  35,928 & 24.3 & \textbf{2.23} \\ 
 \hline
   W4 &  3,364 & 10.8 & \textbf{1.41} \\ 
 \hline
 \end{tabular}
\end{adjustwidth}
\end{table}

However, $\chi^2_N$ is very susceptible to outliers, as well as non-Gaussian, multimodal distributions. Therefore we introduce another method of evaluation which focuses on capturing the match between the data and the full posterior distribution.
A key question is whether or not the posteriors from either SED-fitting code or our empirical model have any overlap with the data. To investigate this, we choose to evaluate the  probability of the data under the posterior

\begin{align}
    \label{eq:average_p}
    \widehat{\mathrm{p}} &= \int \mathrm{p_{data}(WISE)}\mathrm{p_{model}(WISE|optical)dWISE} \\
    & \approx \frac{1}{\mathrm{N_{sample}} }\sum_i^{\mathrm{N_{sample}}} \mathrm{p_{data}(sample_i)}
\end{align}

In short, we evaluate the MC integral of the product of the posterior and data, using samples from the posterior. This can be done directly for \prospector. For \cigale, we treat the posterior as Gaussian, since this is how \cigale~report errors.\footnote{\cigale~has previously existed in a MC version \citep{Serra2011_CIGALEMC}, but it is no longer available. It is possible to calculate a full likelihood landscape, but this can only be done for an extremely small set of galaxies due to memory constraints. For this sample, we have verified that the choice of assuming that the posterior is Gaussian does not constitute a significant error.} For the empirical model, we treat the predictions as delta function posteriors. We approximate $\mathrm{p_{data}}$ as a Gaussian distribution according to the WISE uncertainties. As a baseline, we also evaluate the probability of the data under the data. 

The results are summarized in Table \ref{tab:mean_p_fit}. Figure \ref{fig:PP_plot} shows a so-called PP plot, where each axis show the cumulative probability of the distribution of $\widehat{\mathrm{p}}$. $\widehat{\mathrm{p}}$ is insensitive to outliers, since any catastrophic failure will be bounded by 0. We can observe that both SED-fitting methods produce overconfident and biased results, whereas our empirical method is essentially unbiased.

\begin{table}[h!]
\caption{Mean probabilities of the data under the posterior. Full CDFs are shown in Figure \ref{fig:PP_plot}. The empirical and data distributions are a close match. This statistic is less susceptible to catastrophic errors than the $\chi^2_N$. Bolded values highlight the best-performing model for each band.}
\label{tab:mean_p_fit}
\centering
 \begin{tabular}{| c | c | c | c | c|} 
 \hline
  & \prospector & \cigale & Empirical & Data \\ 
 \hline
  W1 &  0.038 & 0.069 & \textbf{0.66} & 0.70 \\ 
 \hline
   W2 &  0.015 & 0.22 & \textbf{0.70} & 0.70 \\ 
 \hline
   W3 &  0.07 & 0.12 & \textbf{0.63} & 0.70 \\ 
 \hline
   W4 &  0.12 & 0.21 & \textbf{0.70} & 0.70 \\ 
 \hline
 \end{tabular}
\end{table}
There are many other possible metrics of interest for regression tasks like the one presented here (see e.g., \cite{Jespersen2022_mangrove}), but they all confirm the same picture and are thus not shown.

\begin{figure*}
  \centering
  \includegraphics[width=0.95\linewidth]{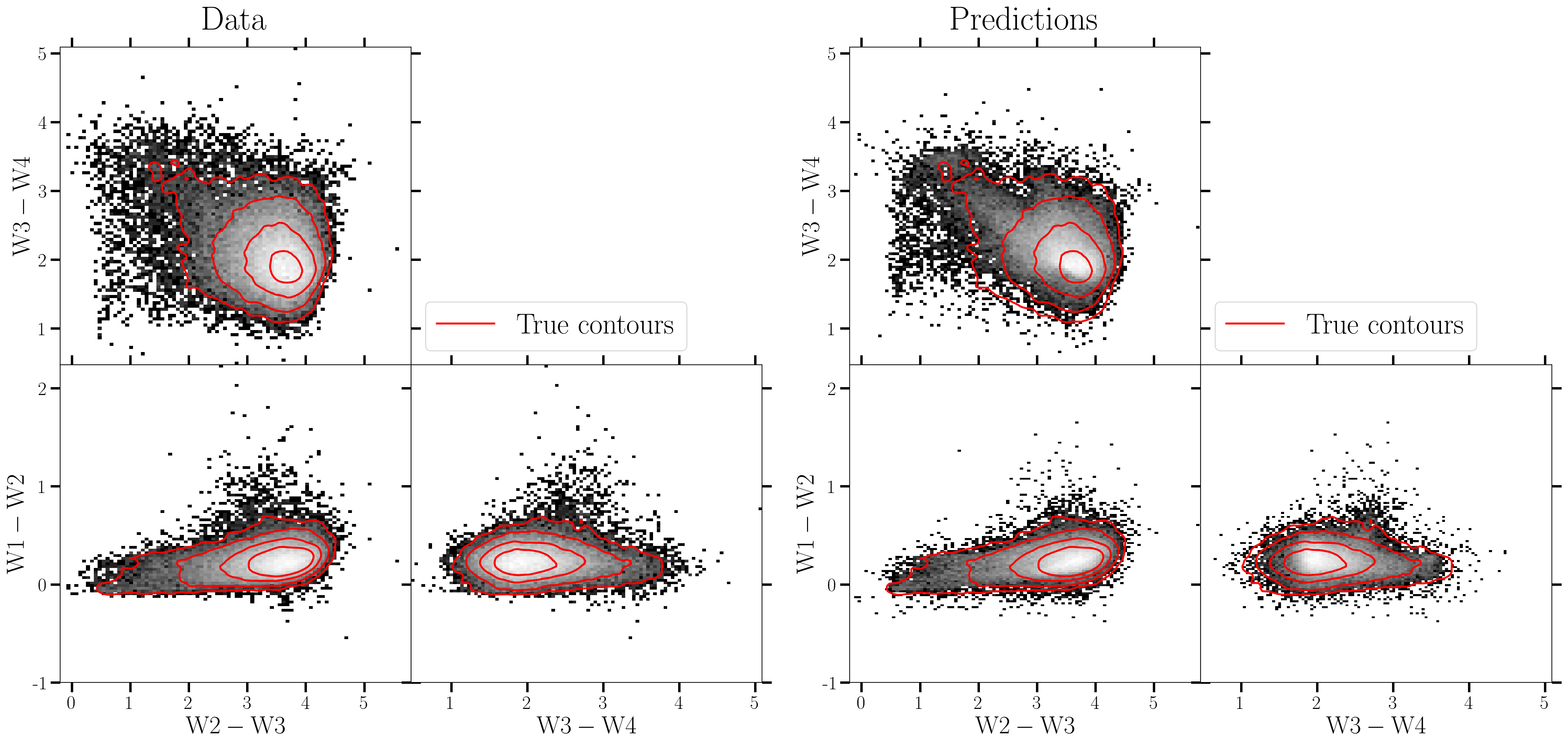}
  \caption{Color-color as measured from WISE (left) and the predictions of our empirical model (right). All data shown are from the test set. The empirical mapping closely reproduces the observed color-color distributions. Extreme values, especially in the bands with large errors (W3 and W4), are disfavoured by the model because they are likely to be produced by measurement error. However, very high values of W1-W2 are also not perfectly represented, likely due to the limited representation of AGN in our training sample.}
  \label{fig:results_spender_color_color}
\end{figure*}

Another important question is whether our empirical model captures not just the individual bands, but also the relations between them, i.e. the colors. We show the color-color plots for the WISE data as well as the predictions from our empirical model in Figure \ref{fig:results_spender_color_color}. The color-color distributions are generally well reproduced. The predicted distributions have fewer outliers than the data distributions since outliers by definition are statistically disfavored and thus effectively downweighted during training. This especially hurts the AGN population (which tend to found at higher values of W1-W2), but this will be discussed later in \S \ref{subsec:results_breakdown_AGN}.

In summary, the empirical model captures the connection between the optical and IR to the extent the uncertainties in the WISE photometry allows. In contrast, the SED modelling codes do not, with both models getting $\widehat{\mathrm{p}}$ and $\chi^2_N$ values orders of magnitude away from the data. \prospector~is generally worse than \cigale. We note that this difference in performance is likely driven by the softening of the uncertainties necessary for \cigale~ to work at all, which artificially widens the posteriors, rather than \cigale~being inherently more accurate than \prospector.

We emphasize that the results here do not constitute a direct comparison between our empirical model and the SED fitting codes because the former was trained directly on the combination of SDSS spectra and WISE photometry in a supervised fashion, whereas the latter ones have only been tuned to reproduce similar data, but neither the specific galaxy sample nor the specific IR photometry. We therefore do not interpret differences in predictive \emph{precision}. Rather, the combination of \emph{systematic biases and lack of posterior coverage} exhibited by the SED models are indicative of model misspecification, regardless of the existence of any alternative method. Our result present an external validation of SED methods, similar to the validation approach of \cite{Leja2017_prospector_use_1}, which these physical SED models unfortunately fail \emph{despite the existence of the strong panchromatic relations} which our empirical model reveals.

\subsection{Influence of Priors}
\label{subsec:results_prior}

\begin{table}
\caption{The median ratio of the variances of the posteriors and priors, as well as the median KL divergences, for all \prospector~parameters. That either statistic is always solidly greater than 1 shows that data dominates the posterior, not the prior.}
\label{tab:prior_shrink_prospector}
\begin{adjustwidth}{-0.6cm}{}
 \begin{tabular}{| c | c | c | c |} 
 \hline
 Parameter & $\frac{V_{posterior}}{V_{prior}}$ & KL divergence \\ 
 \hline
 $\mathrm{log(Z_{sol})}$ & 1478 & 94.15 \\ 
 \hline
  $\mathrm{dust_{old}}$ & 4951 & 58.21 \\ 
 \hline
  $\mathrm{SFH_0}$ & 7874 & 171.11 \\ 
 \hline
  $\mathrm{SFH_1}$ & 103 & 85.23 \\ 
 \hline
  $\mathrm{SFH_2}$ & 913 & 107.03 \\ 
 \hline
  $\mathrm{SFH_3}$ & 299 & 90.26 \\ 
 \hline
  $\mathrm{SFH_4}$ & 51 & 82.34 \\ 
 \hline 
  $\mathrm{M_*}$ & 20035 & 128.46 \\ 
 \hline 
  $\mathrm{U_{min,dust}}$ & 2.30 & 2.74 \\ 
 \hline 
  $\mathrm{q_{PAH}}$ & 2.29 & 6.15 \\ 
 \hline 
  $\mathrm{\gamma_{dust}}$ & 1.82 & 157.26 \\ 
 \hline 
   $\mathrm{U_{gas}}$ & 447 & 24.54 \\ 
 \hline 
   $\mathrm{f_{AGN}}$ & 5.11 & 3.95 \\ 
 \hline 
   $\mathrm{\tau_{AGN}}$ & 2.50 & 2.25 \\ 
 \hline 
   $\mathrm{\Delta dust_{young}}$ & 3.42 & 22.81 \\ 
 \hline 
   $\mathrm{dust_{index}}$ & 158 & 54.89 \\ 
 \hline 
 \end{tabular}
\end{adjustwidth}
\end{table}

\begin{figure}[ht]
    \centering
    \includegraphics[width=0.95\linewidth]{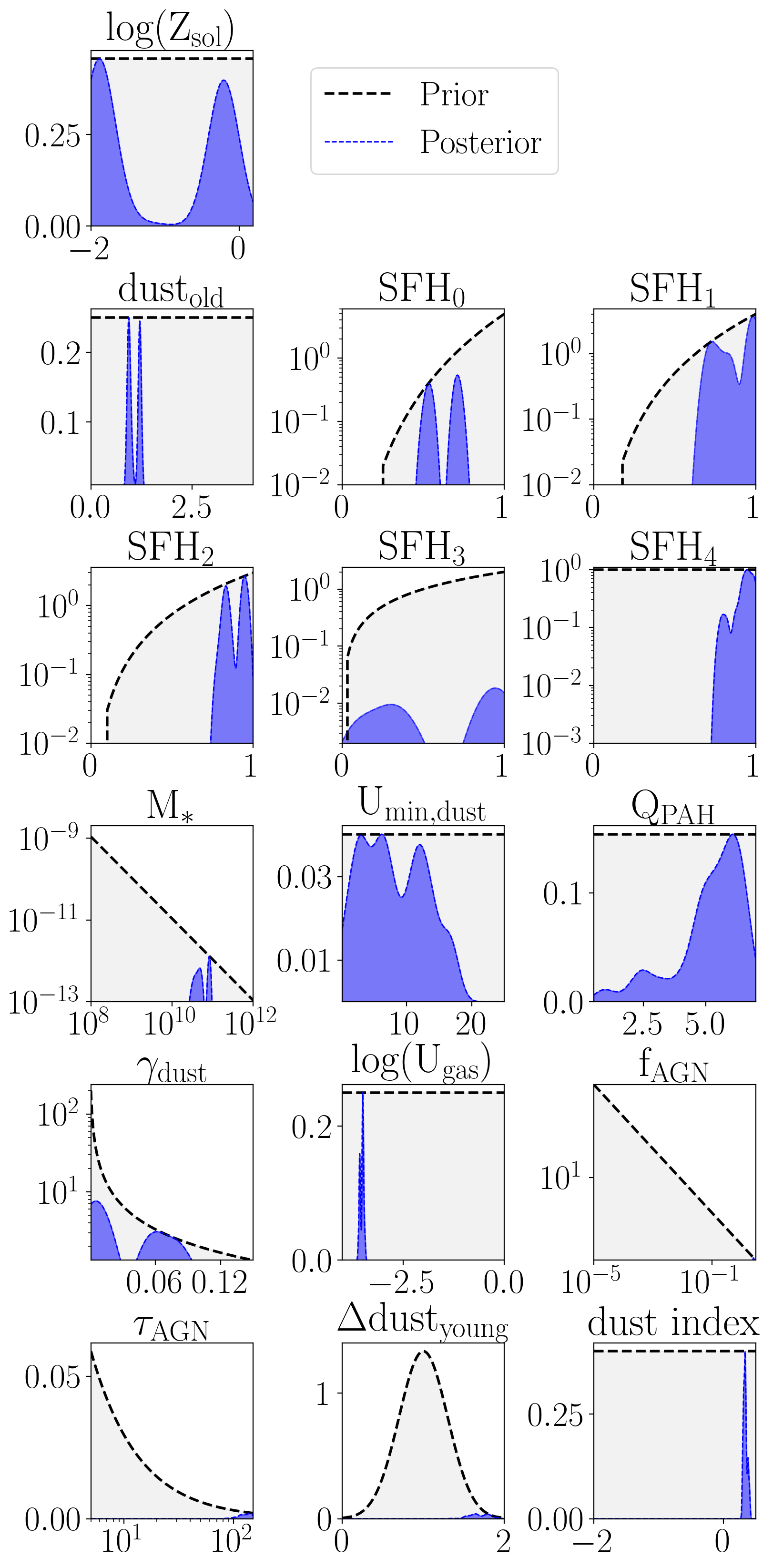}
    \caption{\prospector~priors (grey) and posteriors (blue) for a randomly selected galaxy. The posteriors occupy only a very small part of the prior space. Note that the posterior for $\mathrm{f_{AGN}}$ is barely visible, and concentrated in the bottom right corner. There is no AGN in this galaxy.}
    \label{fig:prior_posterior}
\end{figure}

A central question is to what degree our results are driven by our priors? Below we discuss this for \prospector, \cigale, and our empirical model in turn.

\subsubsection{\prospector}
\label{subsubsec:prospector_prior}
As we can vary the priors in \prospector, we take three different approaches to determine the prior sensitivity. The first is a visual inspection of the priors and posteriors, shown in Figure \ref{fig:prior_posterior}. We see that the ``volume'' of the posterior is much smaller than the prior, indicating that most of the constraining power came from the data, not the prior.

To make the question of whether we are prior-dominated more quantitative, we investigate if the volume of the posterior is significantly smaller than that of the prior by evaluating the ratio of the variances of the posterior and prior. If this ratio is significantly greater than 1, the data dominates the posterior. However, this ratio is insensitive to narrow but multi-modal posteriors, which is why we we also compute the Kullback-Leibler (KL) divergence, defined for two distributions $\mathrm{P}$ and $\mathrm{Q}$ as:
\begin{equation}
    D_{\mathrm{KL}}(P|Q) = \sum_{x\in \mathcal{X}} P(x) \mathrm{log}\left(\frac{P(x)}{Q(x)}\right)
\end{equation}
over a domain $\mathcal{X}$. The KL divergence can be understood as the number of bits of information encoded in one distribution relative to another, and is also known as the \textit{relative entropy} between two distributions.
The variance ratios and KL divergences are summarized in Table \ref{tab:prior_shrink_prospector}, and clearly show that the posteriors are dominated by the data, not the prior.
We must therefore conclude that the model likelihood, not the prior choices, is responsible for the biases in the \prospector~results that can be seen in Figure \ref{fig:PP_plot}.
With the error model of the SDSS spectra being well-described by the Gaussian form employed by \prospector, the problems must originate in the actual SED model and its inability to reproduce the data (as evidenced in Figure \ref{fig:demonstration}).

Lastly, we evaluate if the posterior is concentrated towards the ends of the prior, indicating that our priors may be too narrow. We find that the posterior is usually not concentrated towards the edge of the prior. However, in certain cases, the model drives the posterior to the edge, even without evidence, as is the case for the galaxy shown in Figure \ref{fig:prior_posterior}.

\subsubsection{\cigale}
\label{subsubsec:cigale_prior}
For \cigale, the chosen parameter grid represents an effective prior, and we have verified that expanding/limiting the parameter set does not significantly influence our findings. As an example, when expanding the dust emission parameters, taking $\gamma$ from \{0.1, 0.5, 0.9\} to \{0.01, 0.02, 0.04, 0.1, 0.4, 0.9\}, and $\mathrm{U_{min}}$ from \{0.25, 1.0, 4.0, 17, 40\} to \{0.1, 0.3, 1.0, 2.0, 4.0, 8.0,  17.0, 40.0\} \citep[see e.g.,][]{magdis2012_gamma, Ciesla2014_herschel}, an additional higher value of main stellar population e-folding time, as well as an additional low value of fracAGN \citep[see e.g.,][]{Mountrichas2024e}, results improve by a few percent, with both $\chi^2_N$ getting slightly lower, and the posterior coverage getting marginally better. These marginal improvements mainly result from larger predicted uncertainties on the WISE photometry. Future methods could potentially include even more model components and values, but with the native $\texttt{CIGALE}$ installation, the current parameter grid takes a few $10^4$ CPU hours to run.\footnote{It furthermore requires strongly subsampling the number of models and data points to fit the run on a 196 core HPC cluster node.} Thus, any further expansion would require fundamental changes in the implementation of $\texttt{CIGALE}$.

\subsubsection{Empirical Model}
\label{subsubsec:empirical_prior}

An important question raised by the strong predictive performance of our empirical model is whether it is merely learning an effective prior for galaxies with the given WISE photometry, rather than extracting physically informative structure from the optical data. We address this directly by comparing predictions obtained using the same learning framework but with progressively less informative input representations. If the results were dominated by a learned prior inherited from the training set, similar predictive accuracy would be expected regardless of whether the inputs consist of full optical spectra, emission-line measurements, or broadband photometry.

Instead, we find a clear and systematic dependence of predictive accuracy on the information content of the inputs. Using the full optical spectra yields predictions consistent with the observed WISE photometry at the level of $\chi^2 \simeq 1$, while restricting the inputs to absorption/emission-line measurements degrades the performance to $\chi^2 \simeq 3$. When only broadband photometry is used, the predictive accuracy deteriorates further, reaching $\chi^2 \simeq 10$. Because the learning method and training targets are identical in all cases, these large differences show that our results cannot be attributed to a learned prior. Rather, they demonstrate that the model is exploiting information contained in the detailed spectral features themselves — information that is not fully captured by summary statistics or inferred physical parameters (a more complete analysis of these tests is presented in Appendix~\ref{appsec:predictions_other_inputs}).

\subsection{Galaxy size}
\label{subsec:results_breakdown_size}

\begin{figure}
  \centering
  \includegraphics[width=\linewidth]{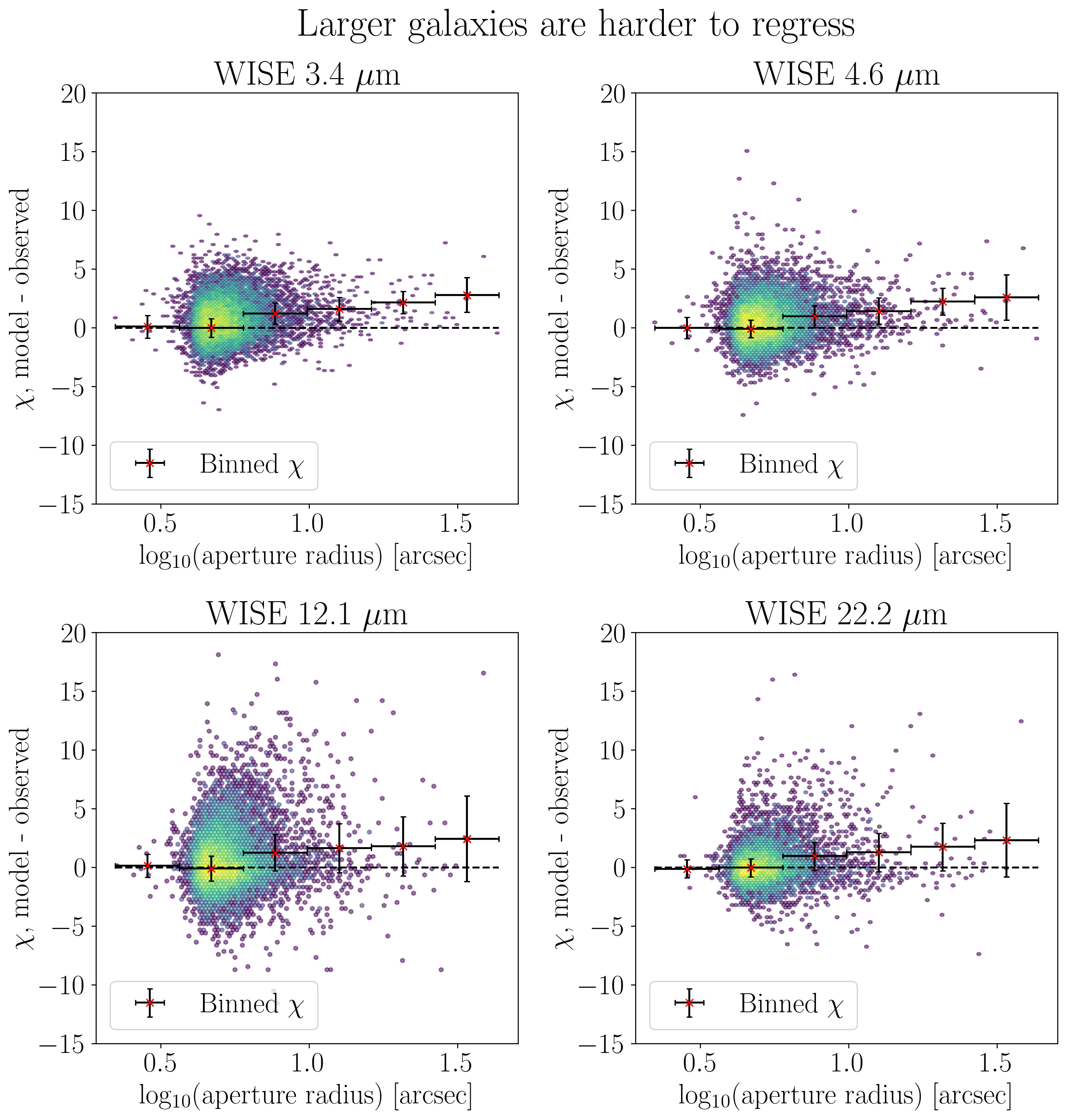}
  \caption{$\chi$ as a function of WISE aperture radius for the predictions from our empirical model, in units of magnitude. Binned trends are plotted for ease of visualization, and show that our model underpredicts the flux (overpredict the magnitude) for larger galaxies.}
  \label{fig:size_fiber}
\end{figure}

One potential caveat in our work is that the WISE photometry and SDSS spectra are measured differently, with the WISE photometry being measured from profile fits inside a variable aperture, and the SDSS spectra being taken in fixed apertures of 3''. Although the spectrograph fiber should contain most of the light from most of the galaxies in the sample (especially for galaxies at $z>0.1$), larger galaxies will inevitably be truncated, which will result in sub-optimal predictions for our models. Figure \ref{fig:size_fiber} shows our prediction accuracy as a function of the size of the boundary used for the WISE profile fit, showing a trend with galaxy size, which we could in principle correct by using aperture-matched photometry. The trend behaves exactly as expected, with our empirical model increasingly underpredicting (overpredicting) the flux (magnitude) for larger galaxies, where the spectrum contains a smaller fraction of the total light of the galaxy. Correcting for this trend brings our $\chi^2_N$ down by $\approx 5\%$. This trend is not necessarily an indication of galaxy size being a physically important parameter, just that the intrinsic spectrum and broadband photometry of a galaxy are determined by the exact stellar population in the measurement aperture.

Based on this analysis, we have attempted to restrict our SED-fitting analysis to galaxies where the SDSS and WISE apertures are better matched (at $z>0.1$). The restriction does not make a significant difference for \textit{predictions}, since the SED predictions are dominated by model misspecification errors. Aperture corrections based on the difference between WISE profile/aperture magnitudes seem to make a small difference for the low-$z$ sample when the SED codes are used to jointly fit the IR and optical. However, since predictions are the focus of this work, we conclude that the aperture correction is a subdominant error.
\subsection{Active Galactic Nuclei}
\label{subsec:results_breakdown_AGN}

\begin{figure*}
  \centering
  \includegraphics[width=0.9\linewidth]{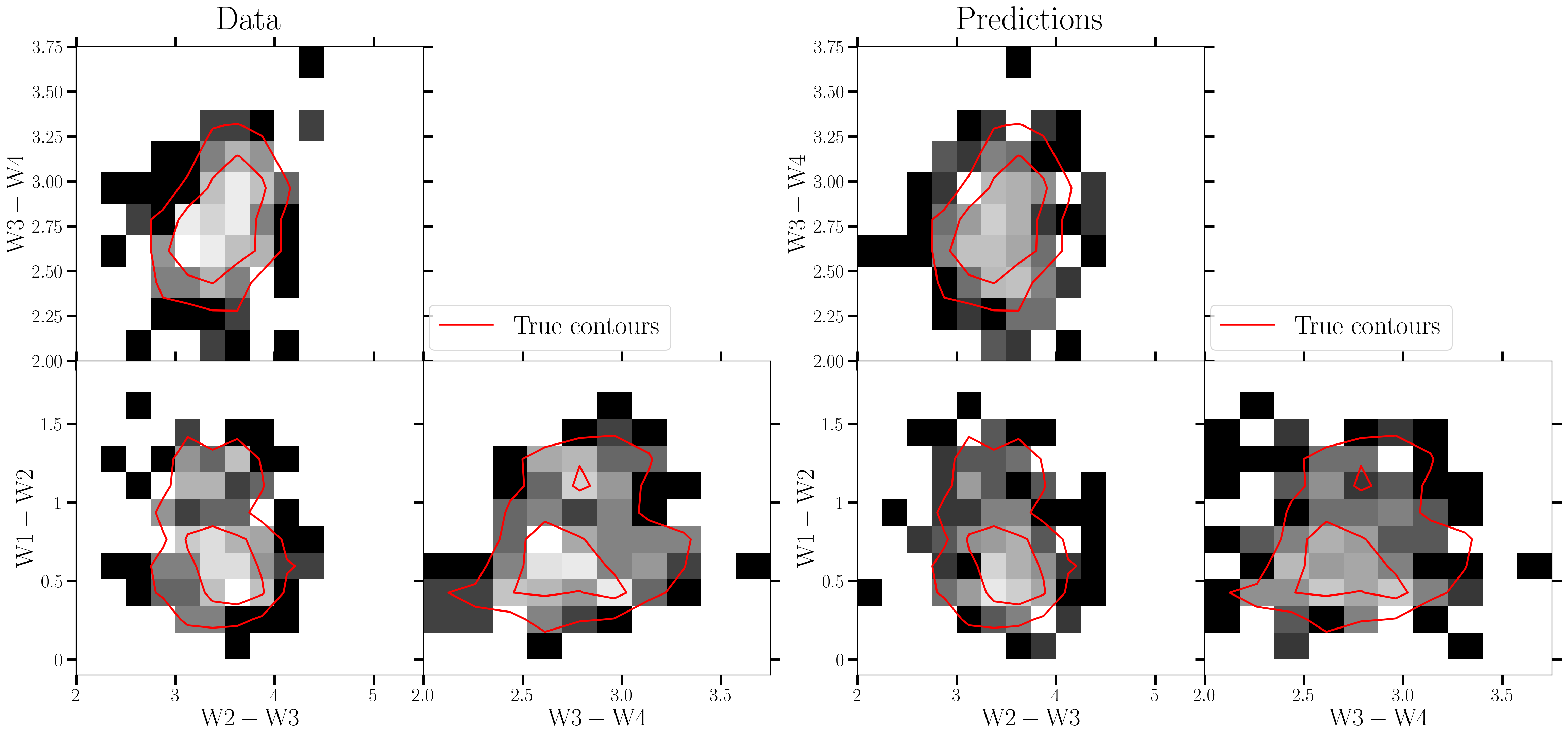}
  \caption{The color-color predictions of our empirical model as shown in Figure \ref{fig:results_spender_color_color}, but only for X-ray selected AGN. Although the results are slightly worse than for the general galaxy sample, the IR emission of AGN is generally predictable from the optical spectrum alone. However, very high W1-W2 values are clearly still disfavored by the empirical model.}
  \label{fig:AGN_precision_color_color}
\end{figure*}

The IR emission from Active Galactic Nuclei (AGN) is often thought to be unpredictable from observations in the optical alone \citep{Nikutta2014_meaning_of_WISE_colours, LyuRieke2018_AGN, Brown2019_agn, Bornancini2022_agn, CruzGoulding2023_modelling_AGN}, and therefore requires special consideration here. To select a sample of high-confidence AGNs, we crossmatch the SDSS spectra with X-ray detections from the fourth XMM-Newton serendipitous source catalogue \citep{Webb2020_XMM-Newton_catalogue}\footnote{Available at \href{https://heasarc.gsfc.nasa.gov/W3Browse/xmm-newton/xmmssc.html}{https://heasarc.gsfc.nasa.gov/W3Browse/xmm-newton/xmmssc.html}} using \texttt{TOPCAT} \citep{Taylor2005_TOPCAT}. We require all sources to be crossmatched to less than 5'' separation, and to be measured as point sources by XMM-NEWTON, since extended sources are likely to be cluster galaxies instead of AGN. To avoid star-forming galaxies, we must implement a luminosity cut, since star-forming galaxies contain many high-mass X-ray binaries which will be detected by XMM-NEWTON. Following \cite{Geda2024_XRB_function}, we set this limit to be $L_\mathrm{XR,AGN}>10^{42} \mathrm{erg/s}$. We use the full 0.2-12 keV flux to calculate the luminosity, applying a K-correction based on a typical photon index of 1.7 \citep{Mushotzky1993_AGN_XR_review}. In Figure \ref{fig:AGN_precision_color_color}, we show the results of predicting WISE photometry for the X-ray selected, test set spectra. Although the results are slightly worse than for the full galaxy sample, there are only weak signs of the model being unable to predict the IR emission of AGN. However, because the AGN hosts generally have high fluxes, and therefore lower relative errors, the color-color plots do not reveal everything. Indeed, if we investigate these results in $\chi$-space, the performance is worse than in non-AGN galaxies, with $\chi^2_N$ being between 2 and 5, depending on the band. Therefore, there must be \textit{some} degree of indeterminability to AGN photometry when one only has access to the optical. This could however also be due to the selection of the SDSS main galaxy sample, which contains only very few AGN. If we were to repeat the exercise for a larger sample of AGN, we would most likely improve the performance substantially. However, our empirical results are still significantly better than the performance of SED models for the same galaxies.

\subsection{Derived physical quantities - AGNs}
\label{subsec:physical_interpretation_discussion_AGN}

\begin{figure*}
  \centering
  \includegraphics[width=0.9\linewidth]{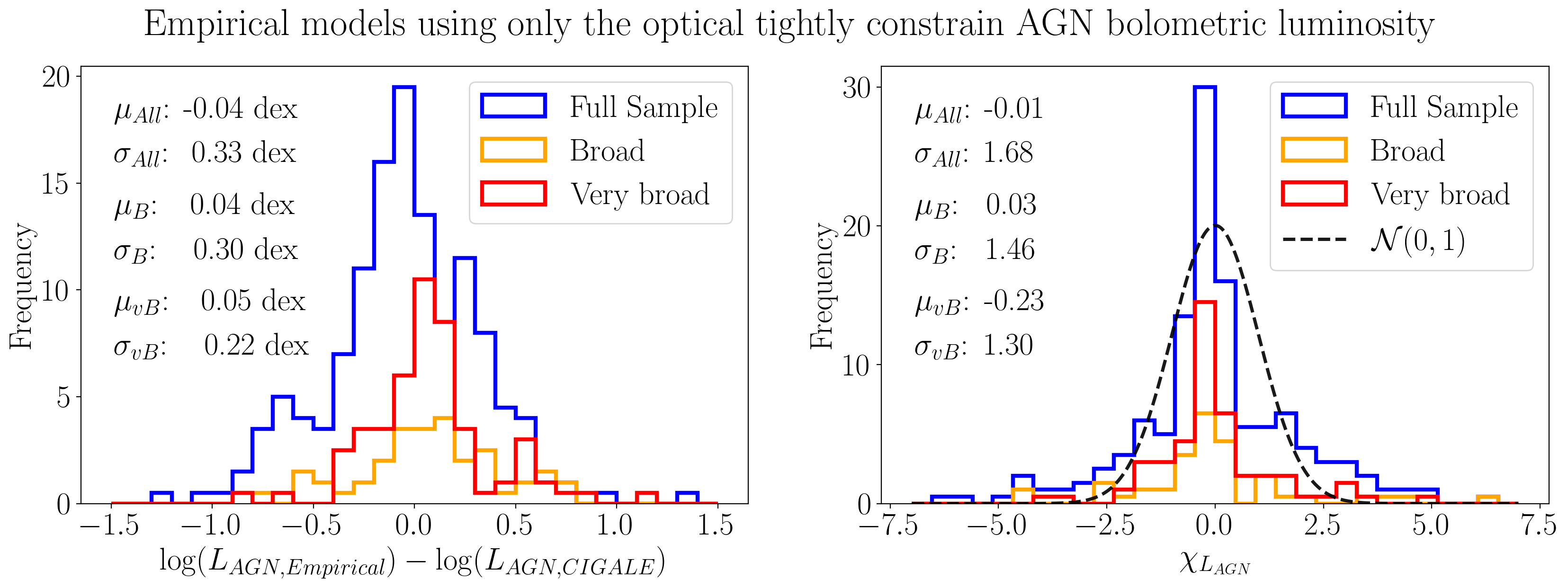}
  \caption{Predictions of $\mathrm{L_{bol}}$ based only on optical information indicate a surprisingly high degree of determinability of physical AGN parameters from the optical. The AGN are X-ray selected, and fitted with \cigale~using optical spectra, $J$, $H$, $K_s$, and WISE photometry. For objects with broad lines (class B, $1000~km/s < v_{\mathrm{FWHM}} < 2000~km/s$), the predictability improves a little, whereas for objects with very broad lines (class vB, $ v_{\mathrm{FWHM}} > 2000~km/s$), the predictability is very high.}
  \label{fig:AGN_Lbol}
\end{figure*}

Our results show that the IR emission of AGN is very well constrained from the optical. This inevitably implies that the physical parameters, such as bolometric luminosity($\mathrm{L_{bol}}$) of the AGN, must also be constrainable. To perform this test, we use the high-confidence, X-ray selected sample of AGN, fit for $\mathrm{L_{bol, AGN}}$ using an \cigale, and then fit another empirical model to learn the relation between the optical and $\mathrm{L_{bol,AGN}}$. In detail:
\begin{enumerate}
  \item The full X-ray selected sample described in \S \ref{subsec:results_breakdown_AGN} is crossmatched with near-infrared ($J$, $H$, and $K_s$ band) data from either the 2MASS survey \citep{Skrutskie2006_2MASS} or from VISTA \citep{Spiniello2019_VEXAS_VISTA, Khramtsov2021_VEXAS_VISTA_2}, depending on what is available.
  \item \cigale~is then used to fit for $\mathrm{L_{bol, AGN}}$, based on the synthetic SDSS, $J$, $H$, $K_s$, and WISE photometry. \cigale~is used with the parameter configuration described in Table \ref{tab:cigale_params}, with the difference that we expand fracAGN to span the [0;1] interval in 50 increments. To make it possible to run \cigale, we limit the remaining parameters to be close to their fiducial best-fit parameters using the original parameter configuration. In general, the fits are good, with $\langle\chi^2_N\rangle = 1.4$, indicating that the fitted physical properties should be informative for further analysis.
  \item A simple MLP, as described in \S \ref{sec:methods}, is then fitted to the outputs of \cigale, based only on the optical spectrum. The performance of the empirical model is then evaluated on a held-out subset (test set) of the sample data.
\end{enumerate}
The results are shown in Figure \ref{fig:AGN_Lbol}. We see that the bolometric luminosity of the AGN can be constrained to within a factor of 2 (0.3 dex) using only information from the optical. This is comparable to the error estimates from fitting for $L_{\mathrm{bol, AGN}}$ using both the optical and IR, as can be seen by the low $\chi^2_N$'s in the right panel of Figure \ref{fig:AGN_Lbol}. The evolution of the performance of our empirical model with optical line width (as shown in Figure \ref{fig:AGN_Lbol}) is also natural; the broader the lines, the easier it is to predict $L_{\mathrm{bol, AGN}}$. The two broad-line samples are defined as having FWHM in either $1000~km/s < v_{\mathrm{FWHM}} < 2000~km/s$ (\textit{broad}) or $ v_{\mathrm{FWHM}} > 2000~km/s$ (\textit{very broad}) for any one of the typical H$\alpha$, H$\beta$, CIV, or MgII emission lines.

In Appendix \ref{appsec:additional_LX_Lpred}, Figure \ref{fig:AGN_appendix}, we show the precision as a function of $\mathrm{L_{bol,AGN}}$, as well as the \cigale~$\mathrm{L_{bol,AGN}}$, and empirical $\mathrm{L_{bol,AGN}}$ predictions as a function of XMM-NEWTON X-ray luminosity.

\subsection{Derived scaling relations - AGNs}
\label{subsec:physical_interpretation_discussion_scaling_AGN}

\begin{figure}
    \centering
    \includegraphics[width=1.0\linewidth]{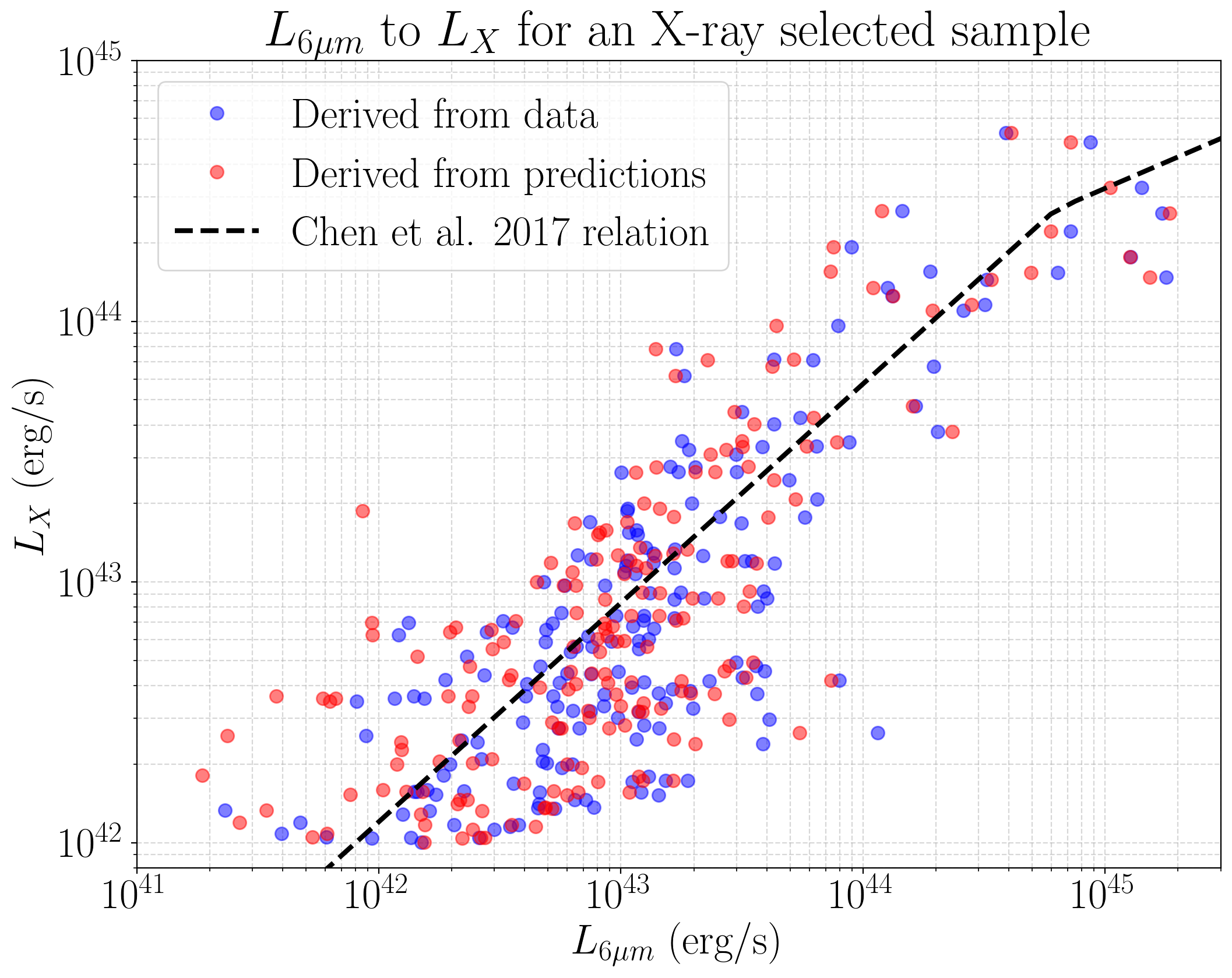}
    \caption{Correlation between X-ray luminosity ($L_{X}$) and 6$\mu m$ luminosity ($L_{6 \mu m}$) found using either real data or the predicted photometry. Both sets of $L_{6 \mu m}$ closely follow the \cite{Chen2017_AGN_XR_IR} XR-MIR relation for AGNs. There is a small bias towards lower $L_{6 \mu m}$ for the predicted WISE photometry, but it is only on the order of 0.06 dex.}
    \label{fig:6um_to_XR}
\end{figure}

We can also investigate if fundamental AGN scaling relations can be reproduced using the predicted photometry. Here we investigate whether the predicted AGN photometry reproduces the X-ray - mid-IR relation found by \cite{Chen2017_AGN_XR_IR} between X-ray luminosity ($L_{X}$) and 6$\mu m$ luminosity ($L_{6 \mu m}$). The 6$\mu m$ luminosities are determined by fitting a power-law to the four WISE bands, and then calculating the \textit{rest-frame} 6$\mu m$ luminosity from the fit. This procedure is done for both the real WISE photometry and the empirically predicted photometry, and the results are shown in Figure \ref{fig:6um_to_XR}. The \cite{Chen2017_AGN_XR_IR} relation is clearly well reproduced by both the data and the predicted WISE photometry. There is a small bias towards lower $L_{6 \mu m}$ for the predicted photometry, especially at low $L_X$. The bias is, however, only around 0.06 dex, and does not significantly change the global scaling relation.
\subsection{Derived physical quantities - PAHs}
\label{subsec:physical_interpretation_discussion_qpah}

\begin{figure}
  \centering
  \includegraphics[trim={0.2cm 0.3cm 0cm 0.cm}, clip, width=\linewidth]{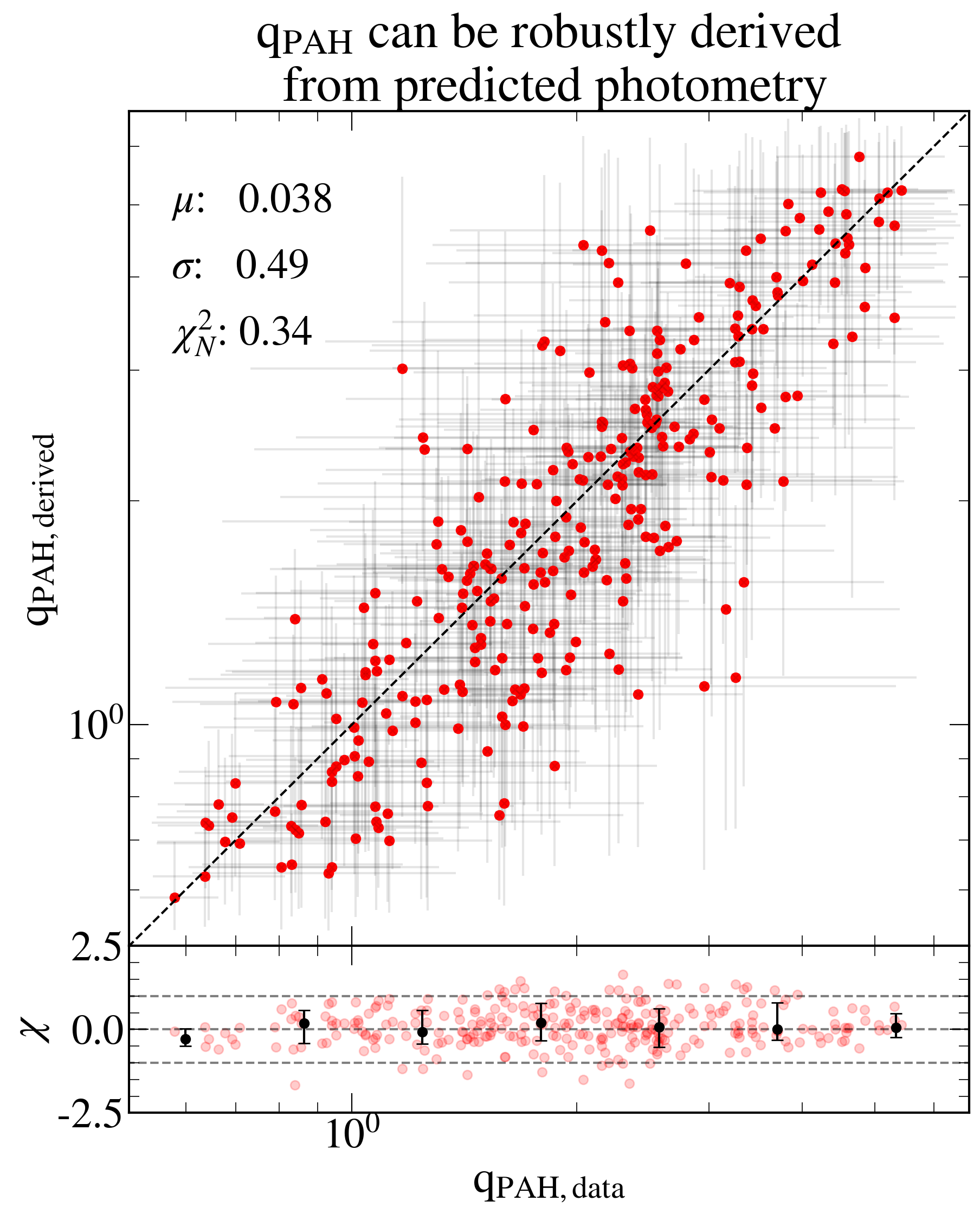}
  \caption{$\mathrm{q_{PAH}}$ estimates from real data and predicted photometry closely agree. Information about PAHs must generally be inferrable from the optical.}
  \label{fig:PAH_constraint}
\end{figure}

Another property, namely the amount and properties of polycyclic aromatic hydrocarbons \citep[PAHs,][]{Draine2007_PAH_SINGS}, is also thought to be constrainable only with IR information. Once again, we rely on a model, this time the \prospector/\fsps~implementation with the \cite{Draine_Li_2007_PAH} dust model, to determine the fraction of dust in PAHs, $\mathrm{q_{PAH}}$. PAH emission strength is a combination of the amount of PAHs ($\mathrm{q_{PAH}}$) and the heating of those PAHs by young stars. To investigate the degree to which $\mathrm{q_{PAH}}$ can be constrained, we must therefore also accurately know the star formation rate. To avoid issues with dust-obscured star formation, we fit $\mathrm{q_{PAH}}$ using WISE photometry alongside photometry from the far-infrared Herschel H-ATLAS survey \citep{Valiante2016_HATLAS}.\footnote{Downloaded from \href{https://www.h-atlas.org/public-data/download}{https://www.h-atlas.org/public-data/download}.} We then repeat the process using the predicted photometry, the results of which are shown in Figure \ref{fig:PAH_constraint}. The estimates closely agree. Considering the range of $\mathrm{q_{PAH}}$, the typical change between the derived and true $\mathrm{q_{PAH}}$ corresponds to a relative error of $\approx7\%$. These constraints are significantly better than the ones obtained by exploiting the normal PAH--metallicity relation \citep{Chastenet2023_PAH_metallicity_PHANGS}. 

\section{Discussion}
\label{sec:discussion}

We have established stable, data-driven mappings from optical spectra to the WISE photometry of SDSS galaxies, which enables us to predict fluxes and IR-derived quantities for a held-out sample down to the noise level. Once these mappings have been determined, there is little significant additional information in WISE photometry that is not already contained in optical spectroscopy. The existence of correlations between optical and IR emission has been known for a long time \citep{RiekeLebofsky1979_IR_review}, but the strength and robustness of the mapping is striking, with near-noise-level prediction at the individual-galaxy level. This result can seem puzzling, given that these two wavelength regimes are dominated by different physical processes. Therefore, the underlying physical processes must be so tightly coupled that galaxies occupy a very small subspace of the full space of possible physical configurations. Some of these subspaces are already known, like the fundamental plane or star-forming main sequence \citep{FaberJackson1976, DjorgovskiDavis1987_fundamental_plane, Speagle2014_SFMS}. Our results suggest that a broader set of galaxy properties, even AGN activity to some extent, must form a narrow fundamental hyperplane. Galaxies must then evolve and move on this narrow hyperplane in ways which can be fully determined from optical spectra.  

\subsection{Implication of correlated wavelength ranges}
\label{subsec:corr_lambda_corr_physics}

To interpret our findings, we consider this matrix of binary options:

\begin{minipage}{\linewidth}
   \centering
  \begin{tblr}{
    colspec = {c|cc},
    row{2} = {lightred},
    column{3} = {lightgreen},
    cell{2}{3} = {lightyellow}
  }
   &   $\lambda_{\mathrm{uncorrelated}}$  &  $\lambda_{\mathrm{correlated}}$ \\
    \hline
    $\phi_{\mathrm{separate}}$ &  & \xmark \\
    $\phi_{\mathrm{connected}}$ &  & \cmark \\
  \end{tblr}
\end{minipage}

Wavelength ranges can be correlated or not; and the physical processes in galaxies are either separable or connected. The latter case is implemented by SED-fitting methods with templates or independent parameters (highlighted in red). The results of the previous section clearly show that IR and optical wavelengths are strongly correlated (highlighted in green), so the left column is empirically ruled out and we only need to consider the two remaining cases. 

\begin{figure}[h!]
    \centering
    \includegraphics[width=0.95\linewidth]{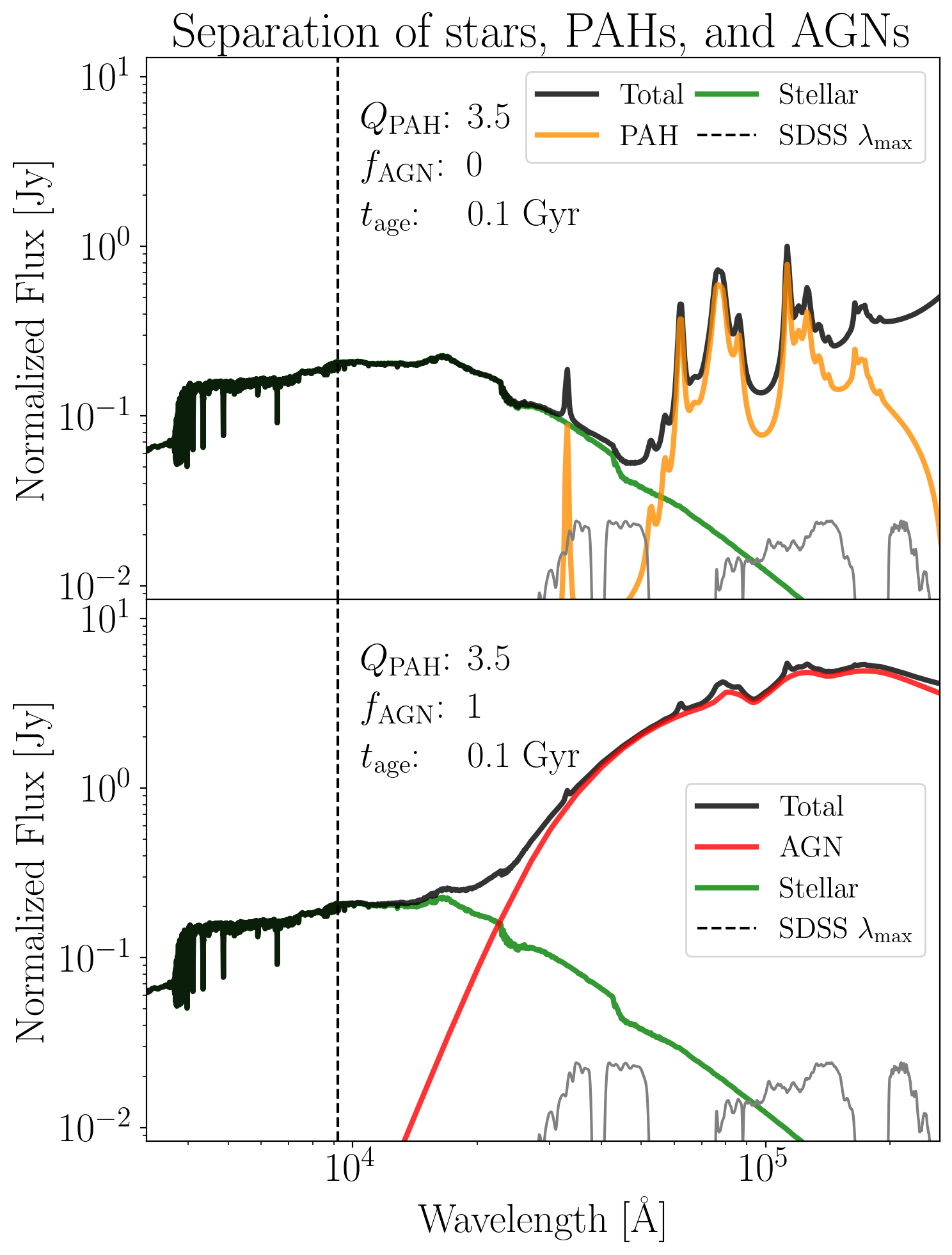}
    \caption{AGN and $\mathrm{q_{PAH}}$ emission dominate the IR, but can be undetectable in the optical. If we can constrain these processes by using information from the optical, there must be a tight underlying coupling between the processes which dominate the optical and those which dominate the IR.}
    \label{fig:wavelength_separation}
\end{figure}

As shown in Figure \ref{fig:wavelength_separation}, the physical processes that dominate the IR can often have no appreciable contribution to the optical. Even at high $\mathrm{f_{AGN}}$, a highly obscuring AGN dust torus will render the optical emission orders of magnitude below the noise limit in the SDSS spectra. If the physics were actually separable, it would imply that there is no way to constrain the IR from the optical. The only remaining option is therefore that the underlying physics are so tightly coupled that the IR is predictable from the optical, even in the case where no AGN emission is discernible in the optical.

We emphasize that this analysis is intended to identify areas where current SED models can be improved, not to diminish their considerable utility. Both \prospector~and \cigale~remain state-of-the-art tools that have enabled a wide range of important scientific results. Our findings highlight where and how the combination of model misspecification and the assumption of separable components leads to biased posteriors, and where data-driven approaches can help guide future model development.

\subsection{Consequences of the separability assumption}
\label{subsec:sed_discussion}

\begin{figure}
  \centering
  \includegraphics[width=\linewidth]{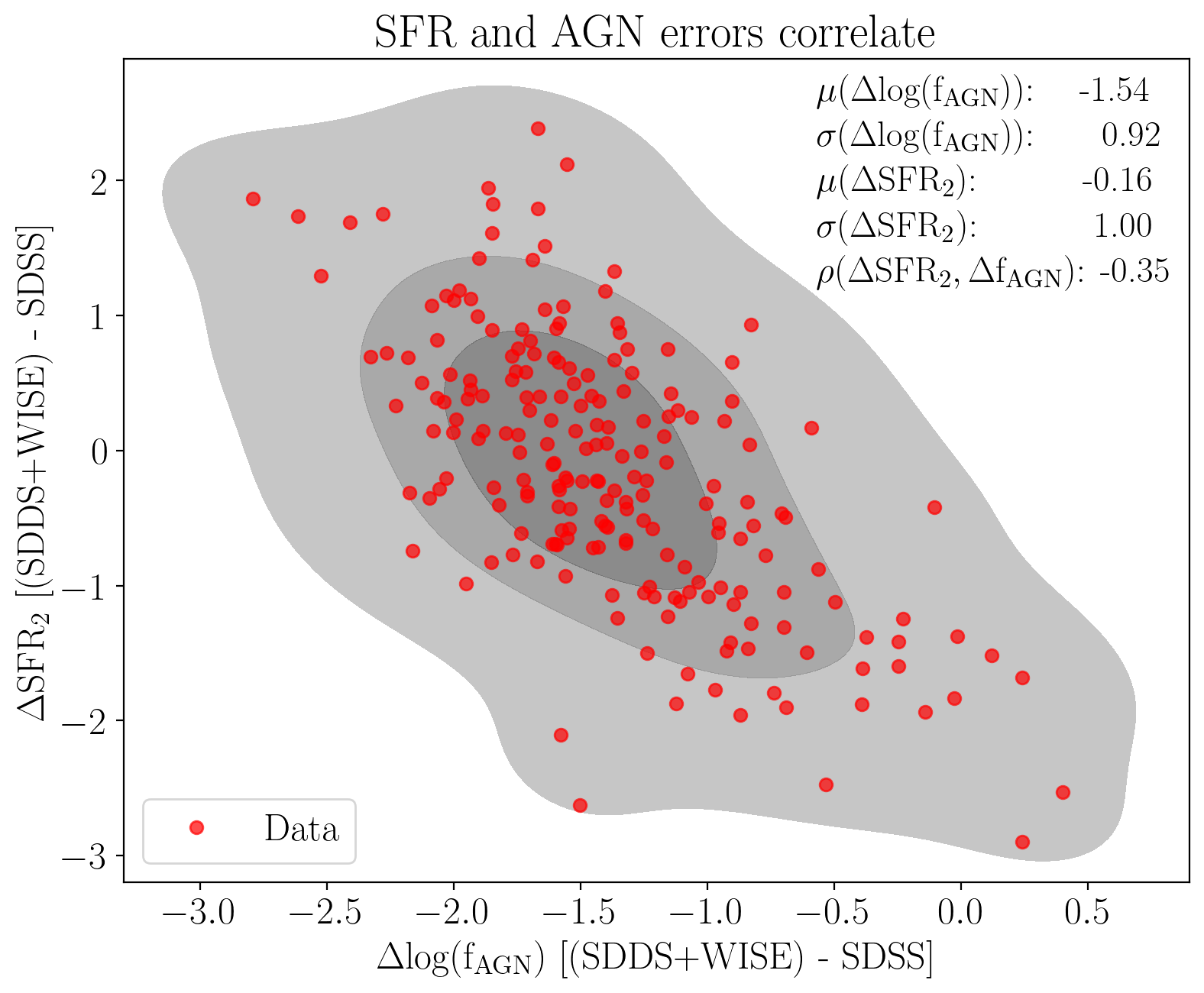}
  \caption{The correlation between changes in \prospector~$f_{AGN}$ (the AGN luminosity as a fraction of the galaxy luminosity) and $\mathrm{SFR_2}$ (SFR on 300 Myr - 1 Gyr timescales) from using the optical only to being constrained by both an SDSS spectrum and WISE photometry. The misestimates are very correlated, meaning that although $\mathrm{SFR_2}$ is not biased on a population-level, any given galaxy could consistently have misestimated $\mathrm{SFR_2}$ on the order of 1 dex. The AGN-SFR degeneracy is known for photometry, but with a spectrum, the degeneracy should break.}
  \label{fig:prospector_fAGN_SFR_errors}
\end{figure}

Although it is now clear that the assumption of separability is wrong, it is tempting to think that as long as one treats the physical components independently, parameter inference may be imprecise but not biased. That is not the case.

As an example, we investigate estimates of star-formation rates (SFRs), one of the main parameters which \prospector~was developed to infer.
Because SFRs are treated independently of $f_{AGN}$, there should be no correlation between these parameters. In particular, misestimating one should not influence the other. Figure \ref{fig:prospector_fAGN_SFR_errors} shows that the wrong treatment of the AGN heavily biases the SFR, correlating the misestimates of $\mathrm{f_{AGN}}$ and the SFR.\footnote{Here we show $SFR_2$, defined as the SFR on 300 Myr - 1 Gyr timescales. The AGN treatment also biases other SFR estimates, but this timescale is the most affected.} 
The resulting biases are commonly on the order of 1 dex. This error is significantly higher than other possible error sources for deriving robust physical properties of the galaxy population from spectra \citep{ConroyGunn2009a_SSPI_uncertain_parameters, Speagle2014_SFMS}. Since it is a systematic error, whose origin is common to most SED-fitting codes, it is especially hard to mitigate. The degeneracy is known for photometry \citep{Ciesla2015_IR_AGN}, but with a spectrum, our results show that the degeneracy should be broken. The biases will become increasingly evident as the JWST MIRI instrument continues to uncover this area of the wavelength space and the associated systematic modelling errors.\footnote{It should be noted that the bias is not a mean bias on the SFR (the mean SFR does not change significantly), but instead strongly biases galaxies conditional on their $\mathrm{\widehat{f_{AGN}}}$.} These biases also motivate current work in developing AGN-specific SED-fitting codes \citep{Buchner2024_AGN_SED}.
 
\subsection{Where is the information coming from?}
\label{subsec:information_lines_discussion}

\begin{figure}
  \centering
  \includegraphics[width=\linewidth]{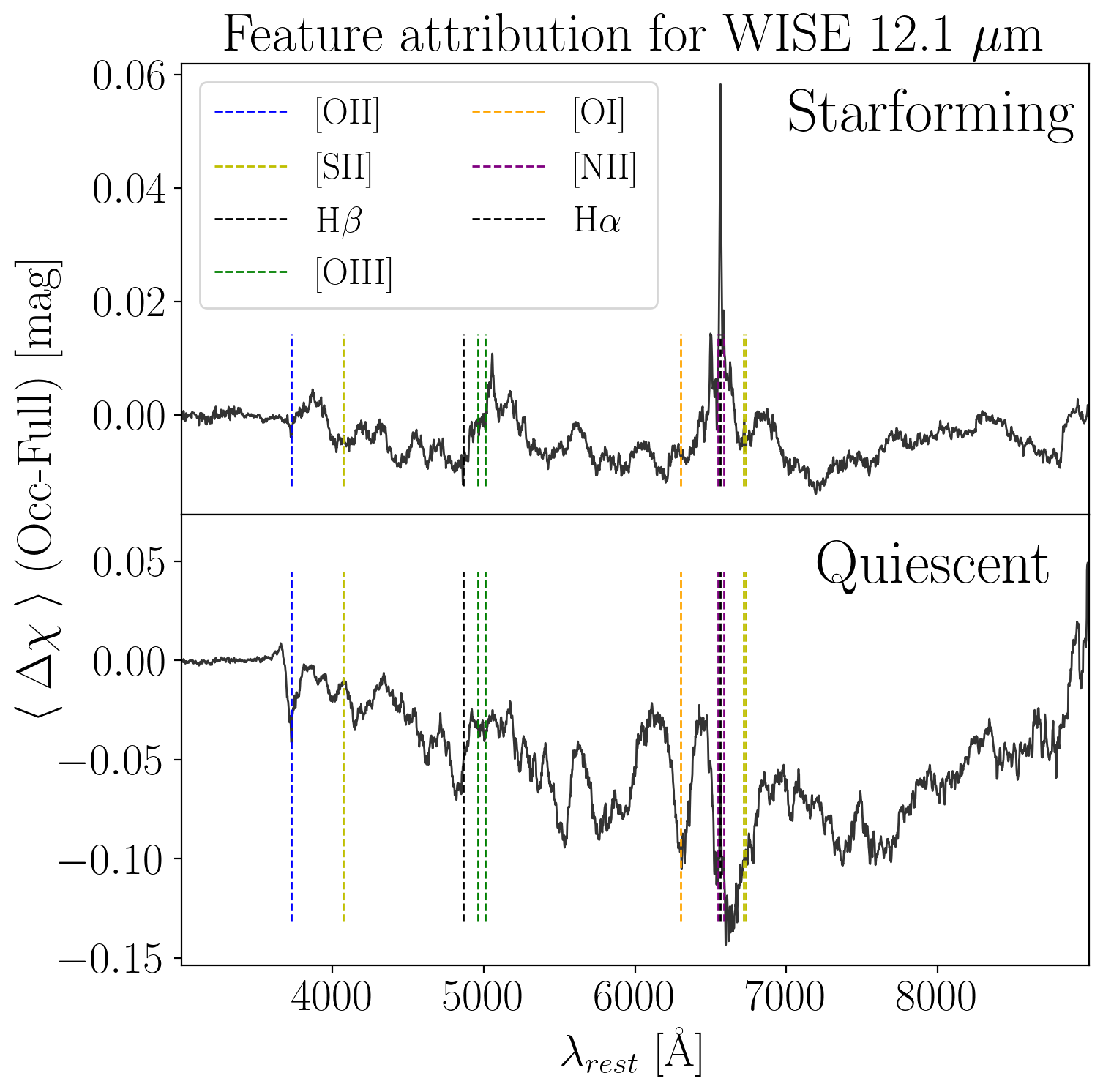}
  \caption{Bias relative to \prospector~calculated with the multiscale occlusion method. Many of the standard ionization lines are extremely important, with H$\alpha$ creating the biggest offset for starforming galaxies, and [OII] creating the sharpest single feature for quiescent galaxies. Together, these lines create an efficient chronometer for each galaxy. For example, a galaxy with no H$\alpha$ but some [OII] can effectively be placed on a post-starburst galaxy track.}
  \label{fig:feature_importance}
\end{figure}

\begin{figure*}
  \centering
  \makebox[\textwidth][c]{ \includegraphics[width=1.1\linewidth]{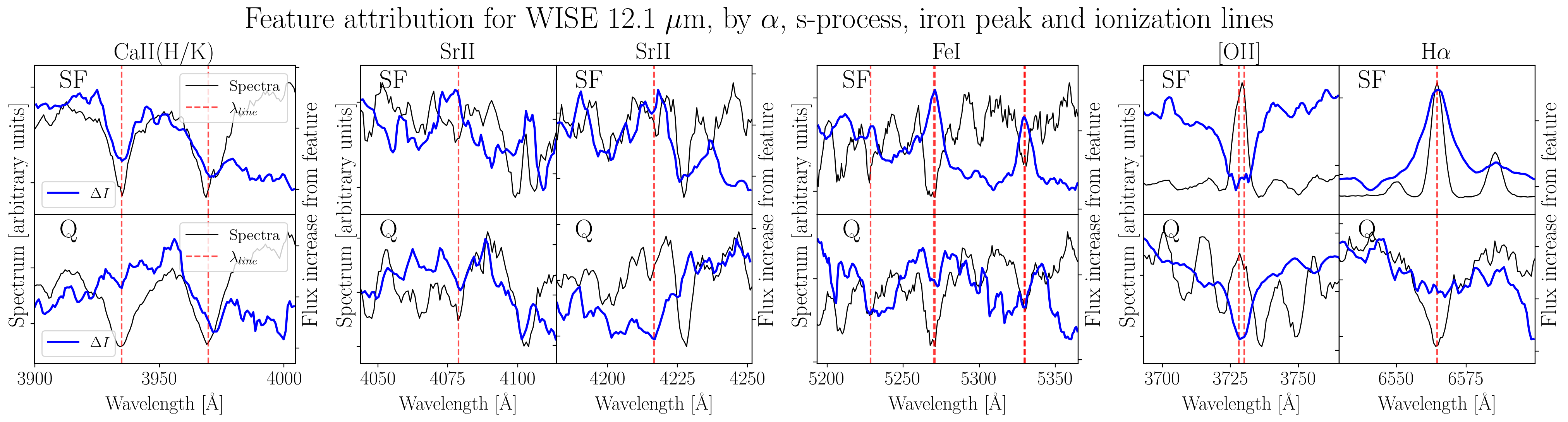} }
  \caption{Feature attribution from multiscale occlusion (blue) overplotted on average spectra (black) for a range of different line systems (red) for the W3 (12.1 $\mu$m) band. The spectra are split into two groups, starforming (top row, SF) and quiescent (bottom row, Q) based on H$\alpha$ strength. The blue feature importance line goes up if an increased strength of the feature in question (deeper absorption lines/brighter emission lines) leads to higher flux in the band. The different line systems have chronologically different origins, with Sr (s-process) being created by star formation on intermediate timescales, Fe (iron-peak) on long timescales and the ionization lines tracing short-term star formation activity. The empirical model places each galaxy precisely on an evolutionary track, with e.g., deeper FeI lines in a starforming galaxy indicating that the galaxy has already undergone a full stellar cycle which has produced the metals needed to create large amounts of PAHs, increasing the flux in the W3 band.}
  \label{fig:feature_importance_zooms}
\end{figure*}

The challenge of self-consistently connecting optical and infrared emission has a long history. The foundational energy-balance argument — that UV/optical light absorbed by dust must be re-radiated in the infrared — has been a cornerstone of galaxy SED modelling since the work of \citep{Calzetti2000, Calzetti2001_energy_balance, Kewley2002_Halpha_IR_SFR} and has motivated composite SFR indicators combining optical and IR tracers \citep{Kennicutt1998_SF_review, KennicuttEvans2012_SF_review}. Our results demonstrate that while this energy balance seems to be broadly respected, the detailed mapping between optical spectral features and IR emission encodes significantly more information than current models capture.

To determine how to improve SED fitting codes, we can now let the data guide us. We use a method called \emph{multiscale occlusion} \citep[MO,][]{Shen2023_multiscale_occlusion}, to determine which features in the spectra allow our empirical model to achieve near-perfect IR predictions.
The method consists of occluding sections of varying lengths with a perturbation and observing the response of our model. We choose to replace sections of the spectrum with Gaussian noise. The noise has the same median flux as the spectral section, but five times the variance. This efficiently occludes any useful features in a given section. We slide an occluding window across the spectrum, replacing each window with noise as described above, encode the perturbed spectrum using \spender, and observe the response of our empirical WISE prediction model. Since we cannot make any \textit{a priori} assumptions about the size of this window, we choose to instead investigate a variety of different window sizes of length N. The results for different N can then be averaged to give an estimate of the average response per pixel across multiple scales. The results are shown in Figure \ref{fig:feature_importance} in the form of $\Delta \chi$, \textit{relative to the initial difference with \prospector}.\footnote{$\Delta \chi = \Delta \chi_{\mathrm{MO}}-\chi_{\prospector}$. A similar figure, but using the $\Delta \chi^2$, can be found in Appendix \ref{appsec:MO}.} The results are split by starforming and quiescent galaxies\footnote{H$\alpha$ is used to classify the galaxies into starforming/quiescent.}, since these two classes have very different responses to the MO. 

In Figure \ref{fig:feature_importance}, many emission lines can immediately be identified, with H$\alpha$ being particularly important, along with other well-known lines from \citet{Kennicutt1992_lines}. However, there is clearly additional information elsewhere. The well-known emission lines marked in Figure \ref{fig:feature_importance} trace the ionization state and short-term star formation activity of a given galaxy, but other properties, such as metal absorption lines, carry important information about star formation on different timescales. 

To further investigate the diversity of time scales traced by different lines, we show a subset of metal lines, as well as zoom-ins on [OII] and H$\alpha$ in Figure \ref{fig:feature_importance_zooms}. The metals analyzed here are $\alpha$ (Ca), s-process (Sr) and iron peak (Fe) elements. The depths of these lines serve as a sensitive chronometer due to the different time-scales of the creation of each element \citep{Conroy2013_strontium_barium}. The fact that these lines all show significant importance is evidence that our empirical model successfully identifies these different timescales. However, not all timescales are equally important for different types of galaxies. For example, since strontium is an s-process element mainly formed in intermediate mass AGB stars, the SrII lines trace star-formation delayed by 200 Myr - 500 Myr \citep{Johnson2020_abundances_post_starburst}. For starforming galaxies, an increased depth of SrII lines means that star formation has already been ongoing for some time, leaving time for AGB stars to produce metals, PAHs, and dust, which are then heated by the ongoing star formation. The combination of increased PAH presence and heating results in stronger W3 emission. For essentially all quiescent galaxies, deeper SrII lines are predictive of lower total W3 flux, since combining deep SrII lines with no H$\alpha$ implies that no star formation has taken place for at least a few hundreds of Myr.

Similarly, Fe forms primarily in Type Ia supernovae, which implies that the FeI lines trace star formation with delays of significantly more than 1 Gyr. For starforming galaxies, deeper FeI lines indicate that the galaxy has already passed through several stellar populations, which have been able to produce the metals that are necessary for forming PAHs \citep{Choban2024_dust_fire}. The combination of metals and high-energy radiation from young stars results in increased W3 flux. Of course, some caution must be taken when interpreting the depth of metal lines, since line depth is not just a function of pure elemental abundance, but also a function of stellar atmospheric parameters, which are harder to interpret. This is especially the case with the CaII H and K lines, although these are still sensitive to the elemental abundance. Note that this improvement is tied specifically to the alpha enrichment rather than overall metal abundance, as all other lines are kept fixed. The details of AGN emission can also be constrained by the ratio between Ca and Fe lines \citep{Martinez2021_CaFe_AGN}, showing their importance.

This analysis shows that although strong ionization lines, like H$\alpha$, are extremely important, models like \prospector~and \cigale~must also get \textit{all} lines from elements up to at least Sr right. Unfortunately, it has recently been shown by \cite{Nersesian2024_SED_models_cannot_recover_lines} that current models \textit{cannot} recover the metal absorption lines correctly under any circumstances.

The analysis further demonstrates that the complex chronology of galaxies \textit{can} be accurately determined from optical spectra. The features which define the theorized fundamental galaxy hyperplane are therefore all present in the optical, which gives important pointers on how to improve our models in the face of the benign indifference of the universe to our modelling preferences \citep{camus1942stranger}.

\section{Conclusion}
\label{sec:conclusion}

In this paper, we establish a strong connection between the optical and infrared emission of galaxies by learning a mapping between optical and WISE photometry. Once learned, the mapping is so strong that our empirical model can predict the WISE photometry of a held-out subset of galaxies down to the noise level ($\chi^2_N \approx 1$) starting from their optical spectra. This builds upon earlier known correlations between optical and IR emission, but the strength of this connection (i.e., lack of scatter beyond instrument noise) is still surprising given that different physical components dominate the emission at these two wavelength ranges. Since a strong connection between optical and IR emission cannot exist without a very strong coupling of the emitting processes, we conclude that those processes must be strongly coupled. This may or may not surprise the reader, but it violates the assumptions made by current SED models.

We demonstrate that although the separability is intuitively enticing, this modelling choice makes SED models unnecessarily biased and overconfident, despite producing good fits in the optical. The key problem we have revealed is not that SED models are less precise; it is that their posterior predictive distributions, derived from the optical spectrum, are inconsistent with the IR data.
The comparison with our empirical model is necessarily asymmetric — our empirical model was trained on large amounts of joint SDSS/IR data — but the diagnosis of overconfident, biased posteriors stands independently of our method and would be apparent from any well-calibrated reference dataset. However, the information needed to predict the IR emission is demonstrably present in the optical. Combined with the issue of overconfident and biased posteriors, our tests reveal shortcomings of ``physical'' (or physically motivated) galaxy SED modelling methods, which rarely incorporate any relations between physical processes, except for energy balance. 

Implementing global physical relations hinges on a detailed understanding of the physics of each component in galaxies and how they interact with others. Fortunately, data-driven models can provide empirical evidence for the most important spectral features, which SED fitting codes should capture but currently do not: common ionization lines, as well as metal absorption lines, provide crucial information about the physical parameters of galaxies, as well as their chronologies. Data-driven models can also help improve our priors, reducing both the runtime and possibly improving inference, as has been suggested by \cite{JiaxuanLi2024_PopSED, Alsing2024_popcosmos} and \cite{Thorp2024_popcosmos}. Another avenue for improving the models would be to incorporate likelihoods which implicitly treat the model misspecifications present in the SED-fitting codes \citep{Czekala2015_flexible_likelihood}.\footnote{Partial code for this kind of likelihood exists in the \prospector~codebase.}

Tests like the one presented in this paper, where the extrapolation abilities of ``physical'' models are tested against data-driven models, can validate the assumptions made in these models. This can also be done for theoretical predictions of galaxy SEDs. As an example, the correlation between wavelengths should be similarly strong in synthetic SEDs produced from MHD simulations such as IllustrisTNG \citep{Pillepich2018_TNG}. This validation of the models can also be done using data-driven models \citep{Woo2024_TNG_spectra}. Another set of next steps concerns using additional datasets, extrapolating in redshift, in spectral resolution, wavelength coverage, and in spatial resolution, all of which will be achieved by current and upcoming surveys \citep{Gunn2012_PFS_cryostats_and_detectors, Takada2014_pfs, desicollaboration2016desiexperimentisciencetargeting, Lang2020_phangs, Casey2023_cosmos-web, PFS_status_2024}. Including more AGN would also be an interesting avenue.

Another interesting direction is to explore a possible connection with latent spaces from other modalities, either other data types like images \citep{WuPeek2020_spectra_from_images, Lanusse2024astroclipcrossmodalfoundationmodel}, or simulated data types. It has recently been shown that empirical representations of the relationships between simulated galaxies and their merger trees or environments can be efficiently learned \citep{Jespersen2022_mangrove, WuJespersen2023_environment, Chuang2024_mergertrees, WuJespersen2024_environment}. Since the inseparability of components implies the existence of a low-dimensional manifold of galaxy properties, one would expect a similar space to exist in these simulation-based representations. Linking these low-dimensional manifolds would provide an indispensable link between theory and data.

\begin{acknowledgements}

The authors wish to thank Rachel S. Somerville, Jenny E. Greene, Sandra M. Faber. Andrew K. Saydjari, Bruce T. Draine, Røbært H. Łåptøn, John F. Wu, Shirley Ho, Lisa J. Kewley, Michael A. Strauss, Matthew Sampson, Vasily Belokurov, Miles Cranmer, and James E. Gunn for valuable discussions before and during this work. The authors also thank Marielle Côté-Gendreau for editing assistance. 
The authors are grateful to Drew Chojnowski for maintaining a useful list of possible galaxy emission lines (\href{http://astronomy.nmsu.edu/drewski/tableofemissionlines.html}{astronomy.nmsu.edu/drewski/tableofemissionlines.html}).

We would furthermore like to thank the two anonymous referees for their reviews. Their perspectives motivated a more careful clarification of the goals and assumptions of our work.

The authors are pleased to acknowledge that the work reported in this paper was substantially performed using the Princeton Research Computing resources at Princeton University which is consortium of groups led by the Princeton Institute for Computational Science and Engineering (PICSciE) and Office of Information Technology’s Research Computing.

This research has made use of NASA's Astrophysics Data System, for which we are grateful.

Funding for SDSS has been provided by the Alfred P. Sloan Foundation, the Participating Institutions, the National Science Foundation, and the U.S. Department of Energy Office of Science. This publication makes use of data products from the Wide-field Infrared Survey Explorer, which is a joint project of UCLA, and JPL/CalTech, funded by the National Aeronautics and Space Administration. This publication makes use of data products from the Two Micron All Sky Survey, which is a joint project of the University of Massachusetts and the IRAC/CalTech, funded by the National Aeronautics and Space Administration and the National Science Foundation. This paper used observations obtained with XMM-Newton, an ESA science mission with instruments and contributions directly funded by ESA Member States and NASA.
This work was supported by the Schmidt Sciences under grant 922.\\

A notebook demonstrating the analysis done in this paper can be found on the first author's website, \href{https://astrockragh.github.io/}{https://astrockragh.github.io/}, or GitHub, \href{https://github.com/astrockragh/IR_optical_demo}{https://github.com/astrockragh/IR\_optical\_demo}. \\

\textit{Software used:} Matplotlib \citep{matplotlib}, pandas \citep{pandas}, sklearn \citep{sklearn}, jupyter \citep{jupyter}, PyTorch \citep{Paszke2019_PyTorch}, numpy \citep{numpy}, AstroPy \citep{2022_Astropy}, accelerate \citep{accelerate}.

\end{acknowledgements}

\bibliographystyle{aasjournal}
\bibliography{ref}

\appendix

\section{Additional figures evaluating the precision of predicting and fitting photometry}
\label{appsec:more_figures_chi2}

Following the failure of the SED fitting models, a question to consider is whether or not the SED models are able to \textit{fit} the photometry when given it. In general, we find that when fitting WISE photometry only, the models are able to perform close to $\chi^2_N = 1$. However, we can next consider the ability of each model to \textit{jointly} fit the IR and optical. The results from doing so are shown in Figure \ref{fig:chi2_N_fit_detections}.

It is clear that while the results greatly improve compared to the prediction case, the reduced chi-square still does not get close to 1, which is what is expected for a good fit. However, as outlined before, when given only the IR photometry, and no optical information, the models can fit the IR photometry to $\chi^2_N \approx 1$, indicating that there is fundamental tension between the optical and IR in the SED models. We have made sure that this is not an issue of normalizing either the spectrum or photometry incorrectly, and the issue persists even when attempting to fit each WISE photometric band individually.

One of the reasons for why the reduced chi-square does not change as much as one may expect is because the optical SDSS spectrum has much higher signal-to-noise, and therefore dominates the likelihood during fitting. 

\begin{figure}
  \centering
  \makebox[0.85\textwidth][c]{ \includegraphics[width=0.8\linewidth]{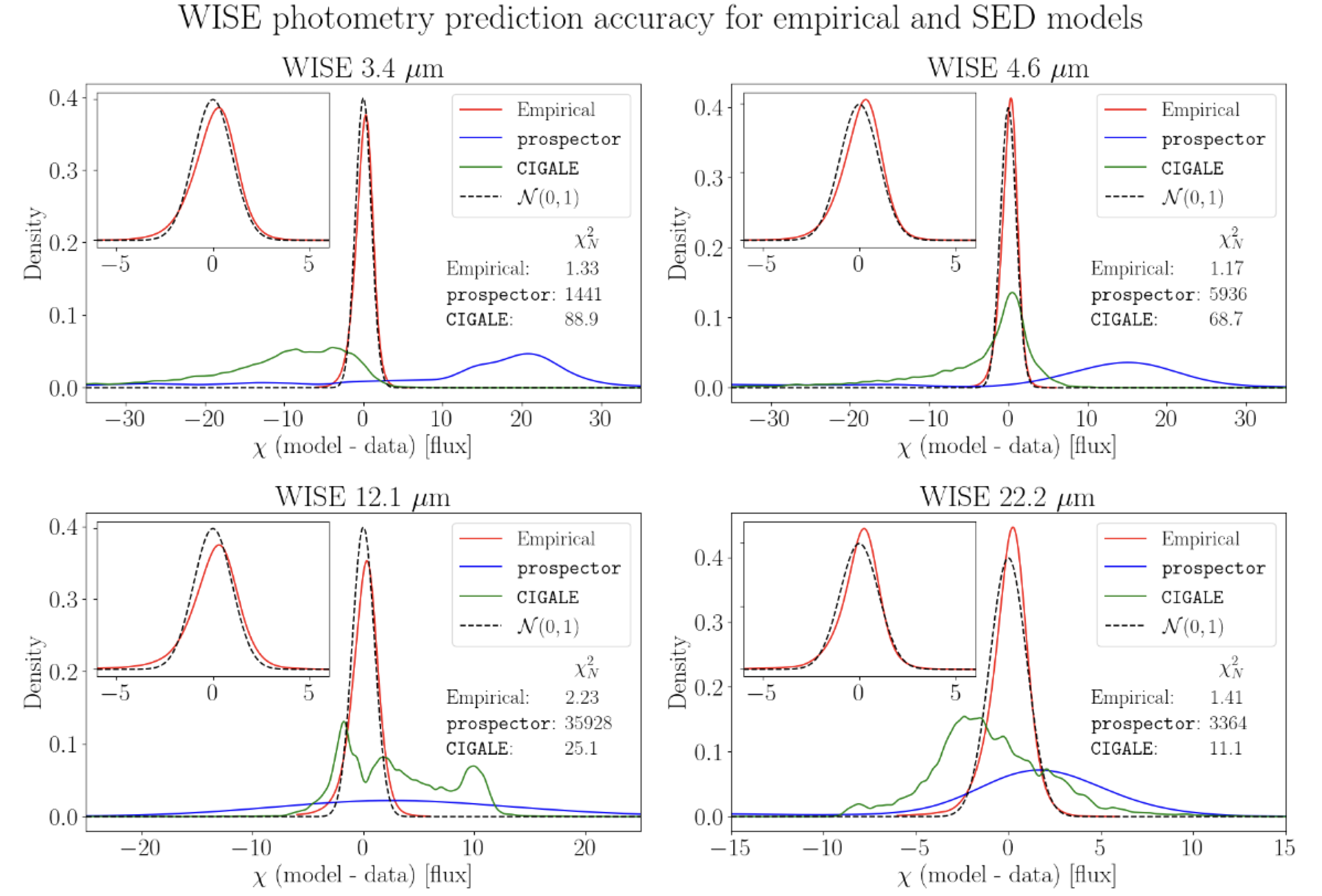} }
  \caption{The $\chi = \frac{F_{model}- F_{data}}{\sigma_{data}}$ distribution for the empirical mapping, \cigale~and \prospector~for all galaxies from the test set with detections in a given band. The empirical model predicts the photometry to $\chi^2_N \approx 1$, indicating a high degree of coupling between the optical and IR emission of all galaxies. The empirical mapping does not produce any extreme outliers, although all galaxy types are included here. \cigale~and \prospector~are both very biased and imprecise compared to our simple empirical model, producing $\chi^2_N$'s in the range of $10^2-10^{4.5}$, many orders of magnitude worse than our simple model. This demonstrates that current SED models cannot correctly capture the coupling of galaxy components. If the modelling were perfect, and the errors Gaussian, the distribution should follow a unit Gaussian, which is included for reference. }
  \label{fig:chi2_N_detections}
\end{figure}

In Figure \ref{fig:chi2_N_detections}, we show the $\chi$-distributions for all galaxies in the test set with detections in a given band. The empirical model captures the connection between the optical and IR to as great of an extent as the uncertainties in the WISE photometry allows. In contrast, the SED modelling codes do not, with both models getting $\chi^2_N$'s orders of magnitude away from unity. \prospector~is generally worse than \cigale. 

\begin{figure}
  \centering
  \makebox[0.85\textwidth][c]{ \includegraphics[width=0.6\linewidth]{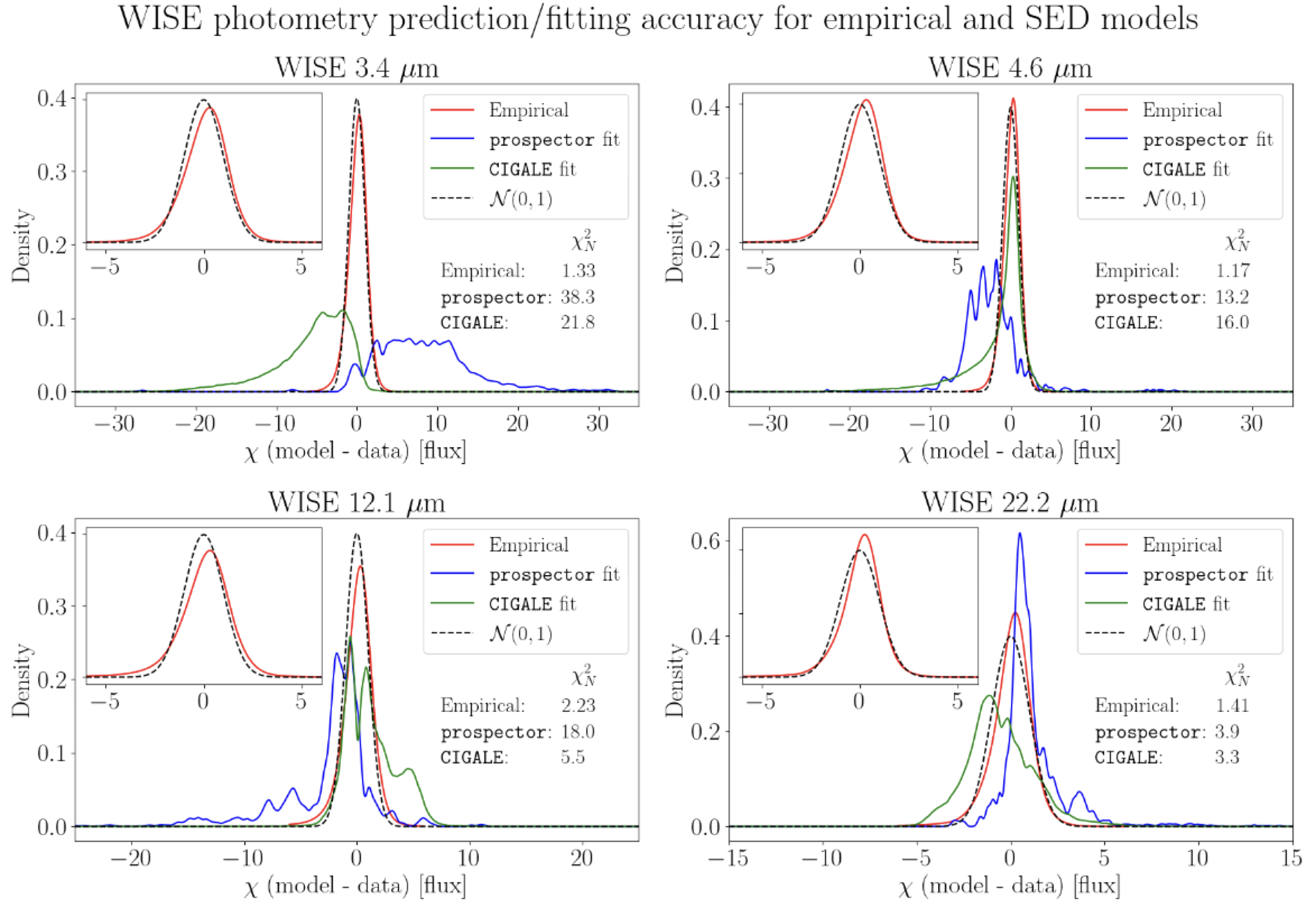} }
  \caption{$\chi^2_N$-distributions for jointly fitting the optical and IR. The $\chi = \frac{F_{model}- F_{data}}{\sigma_{data}}$ distribution for the empirical mapping predictions, as well as fits from \cigale~and \prospector~for all galaxies from the test set with detections in a given band. It is clear that \cigale~and \prospector~continue to be both biased and imprecise compared to the simple empirical model, \textit{even when given the photometry}. However, as shown below, when given only the IR photometry, and no optical information, the models can fit the IR photometry to $\chi^2_N \approx 1$, indicating that there is fundamental tension between the optical and IR in the SED models. }
  \label{fig:chi2_N_fit_detections}
\end{figure}

\begin{figure}
  \centering
  \makebox[0.85\textwidth][c]{ \includegraphics[width=0.6\linewidth]{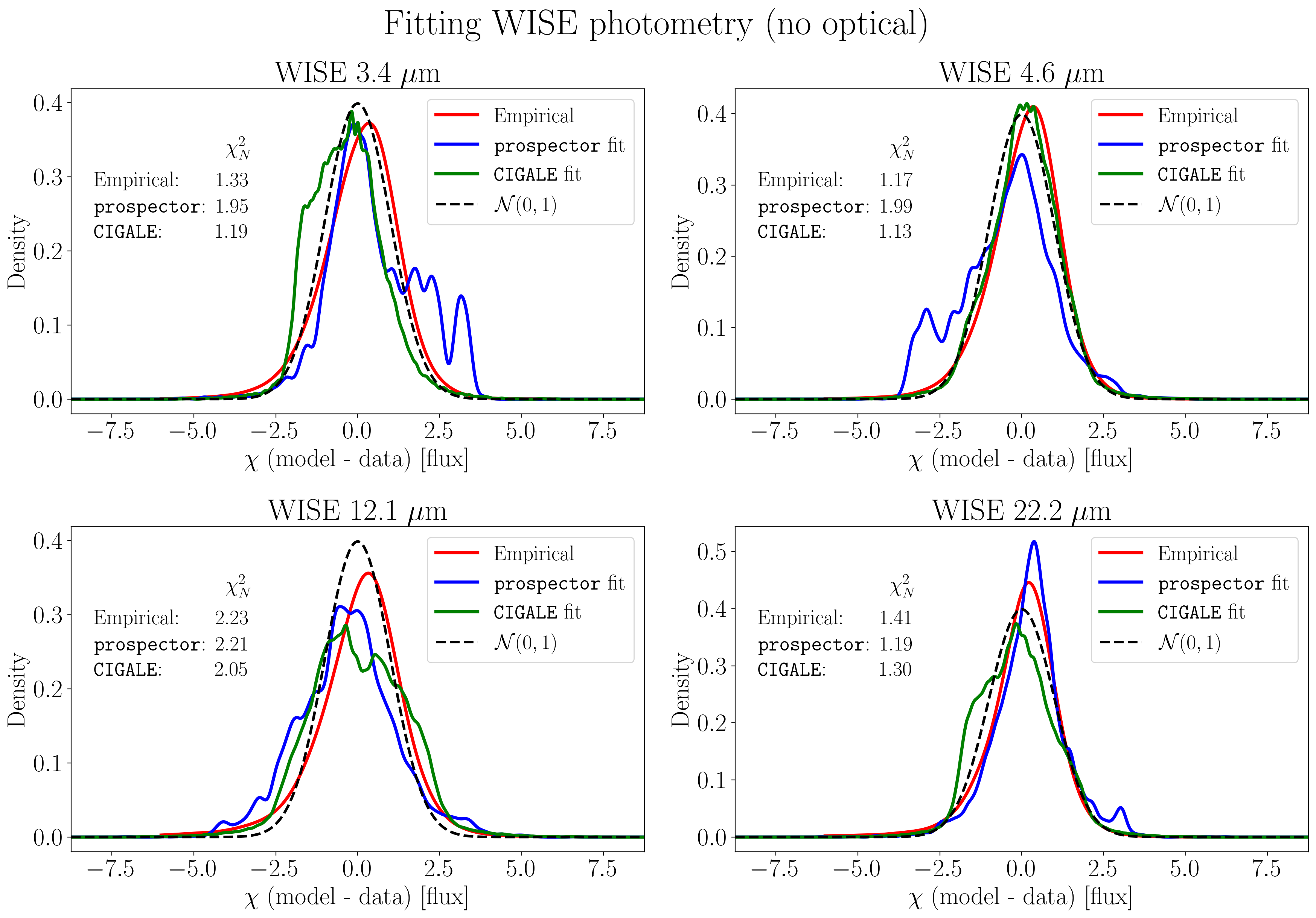} }
  \caption{Same as Figure \ref{fig:chi2_N_detections}, except this is for only fitting the IR. The $\chi = \frac{F_{model}- F_{data}}{\sigma_{data}}$ distribution for the empirical mapping predictions, as well as fits from \cigale~and \prospector~for all galaxies from the test set with detections in a given band. It is clear that without the optical \cigale~and \prospector~do manage to perform similarly to the simple empirical model. However, as shown above, when given both the optical and IR photometry, the models cannot fit the IR photometry to $\chi^2_N \approx 1$, indicating that there is fundamental tension between the optical and IR in the SED models. }
  \label{fig:chi2_N_fit_detections_IR_only}
\end{figure}

\section{Additional precision figures for AGN tasks}
\label{appsec:additional_LX_Lpred}

\begin{figure}
    \centering
    \includegraphics[width=0.32\linewidth]{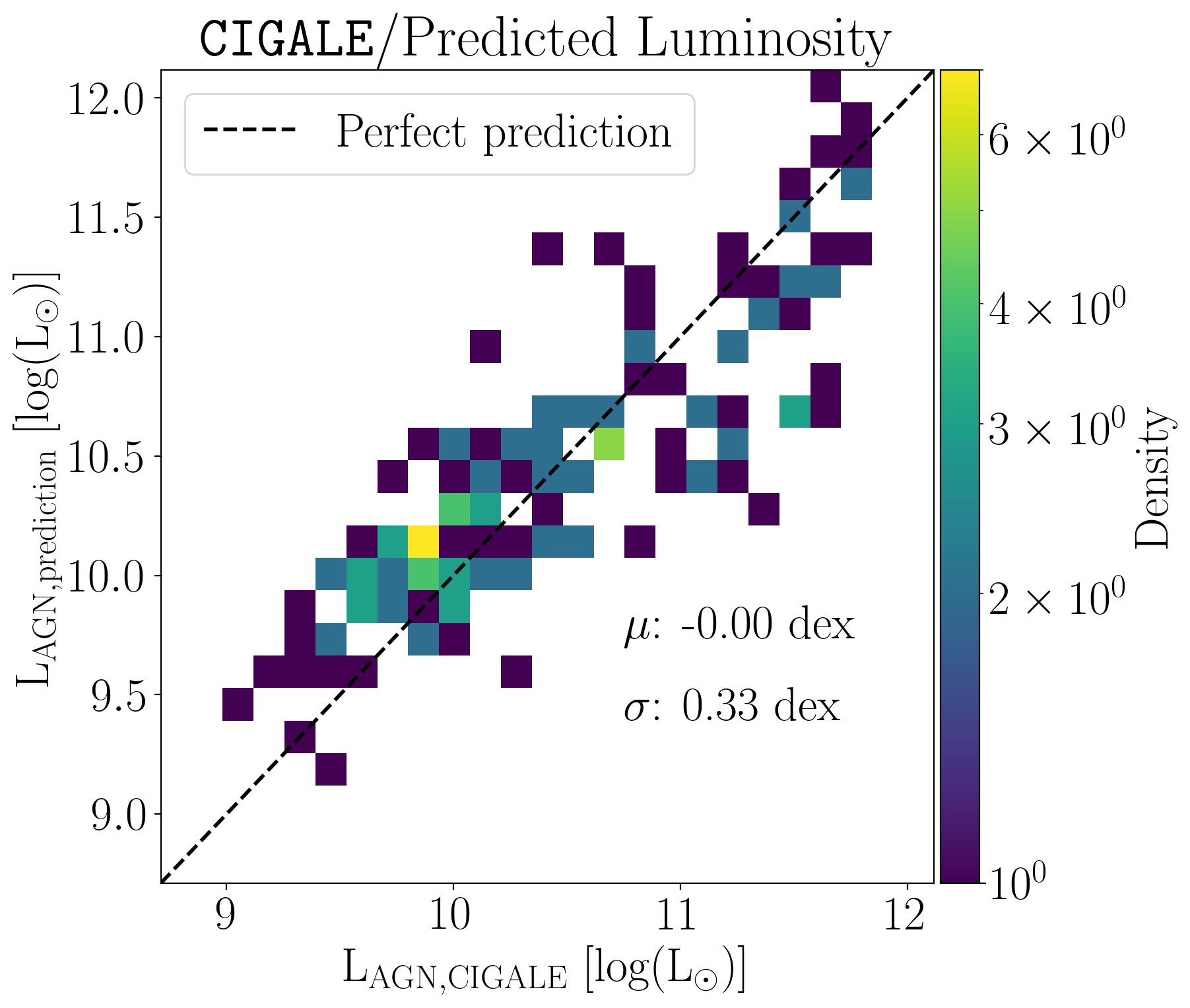}
    \includegraphics[width=0.32\linewidth]{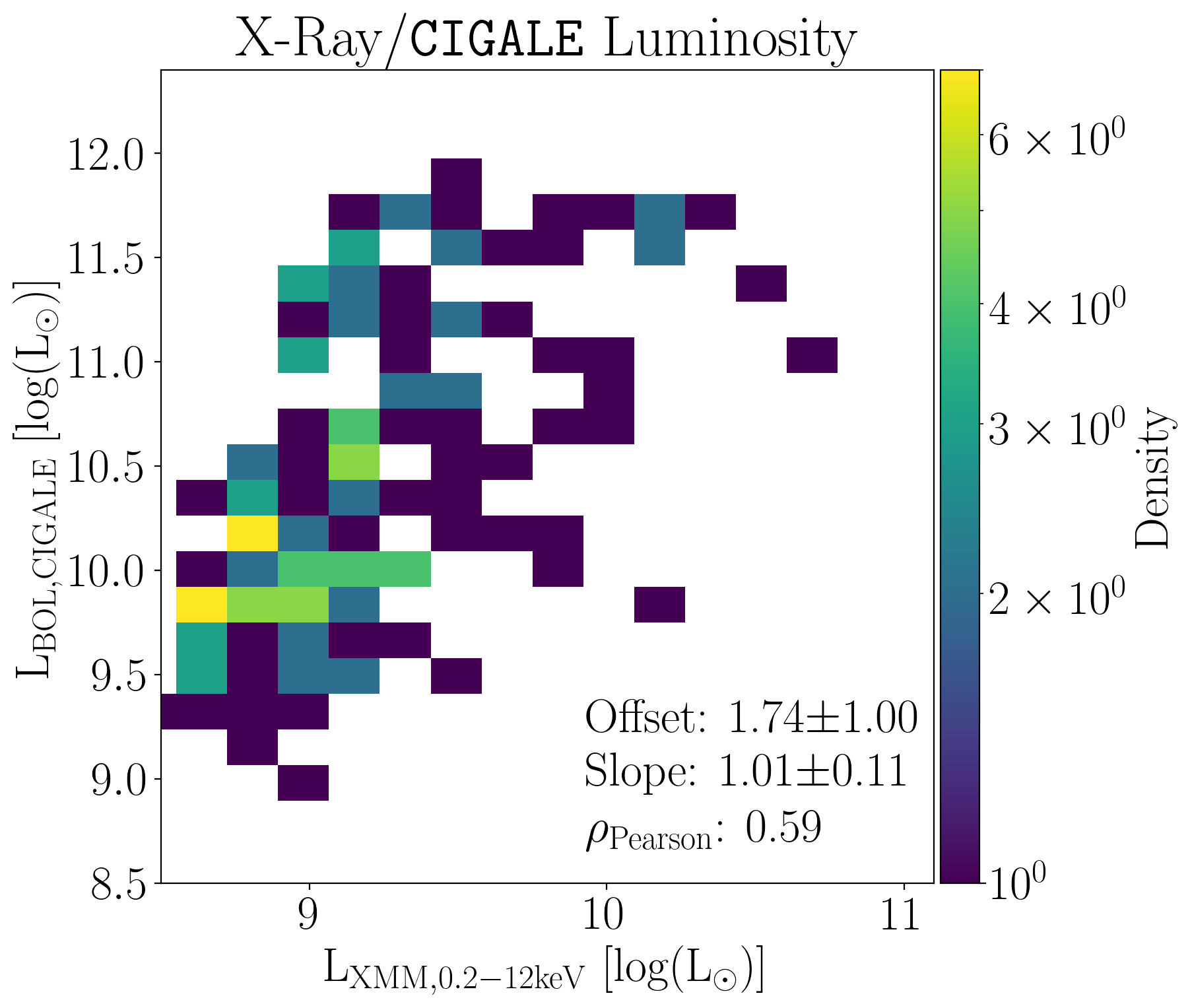}
    \includegraphics[width=0.32\linewidth]{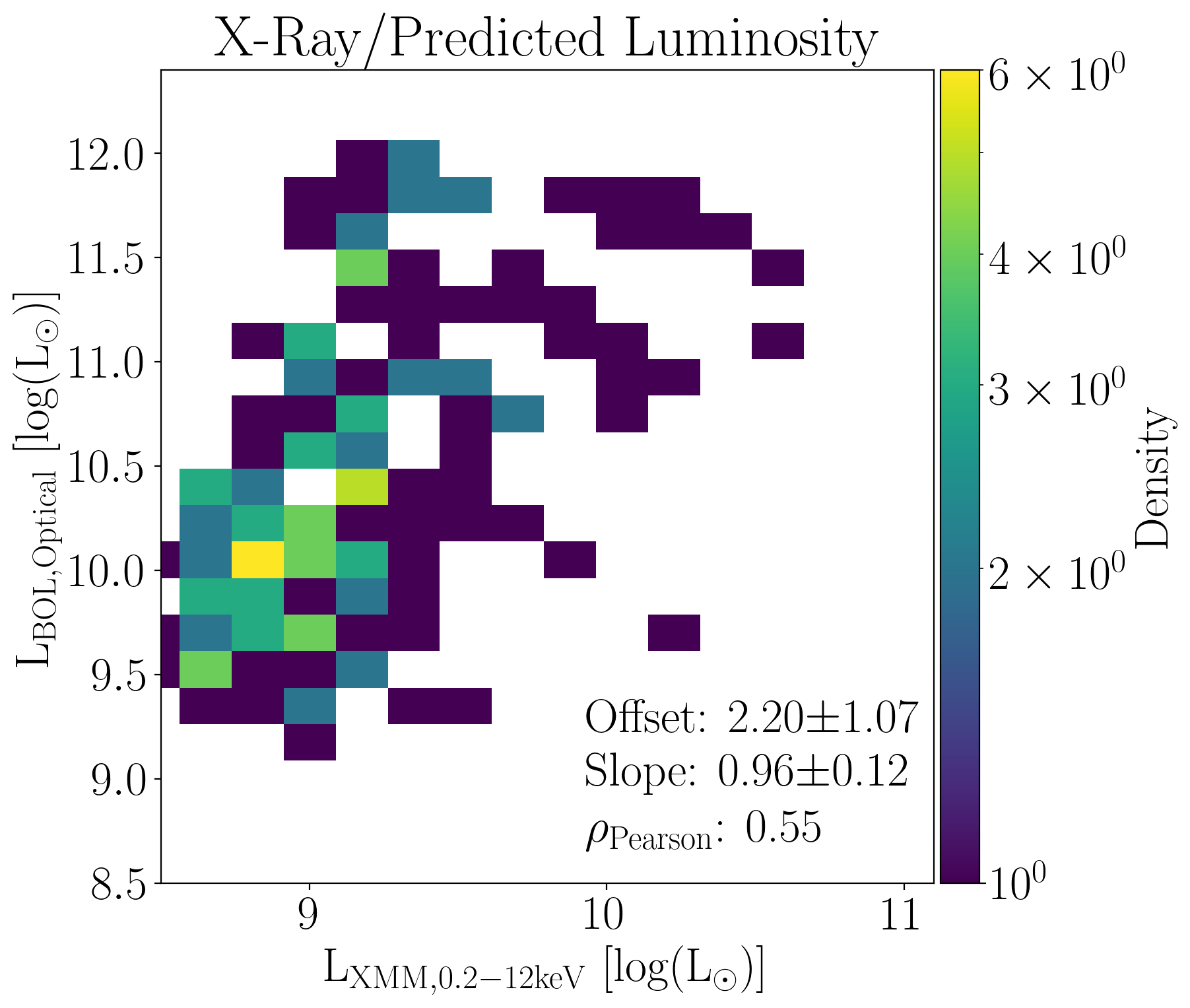}
    \caption{(left) \cigale~fitted bolometric luminosity versus predicted bolometric luminosity. This shows that the predictions are not biased at any specific bolometric luminosity. (middle) XMM X-ray luminosity versus \cigale~fitted bolometric luminosity. Note that the X-ray's are not used for fitting. (right) XMM X-ray luminosity versus predicted bolometric luminosity. The relations are overall the same. All luminosities are given in solar luminosities. These relations are consistent with those of \cite{Spinoglio2024_AGN}.}
    \label{fig:AGN_appendix}
\end{figure}

\section{Additional data information}
\label{appsec:additional_data}

\subsection{Crossmatching WISE, SDSS, and XMM-Newton}
\label{subsec:crossmatch_data}

\begin{figure}[h!]
  \centering
  \includegraphics[trim={0.cm 0.0cm -1cm 0.cm}, clip, width = 0.6\linewidth]{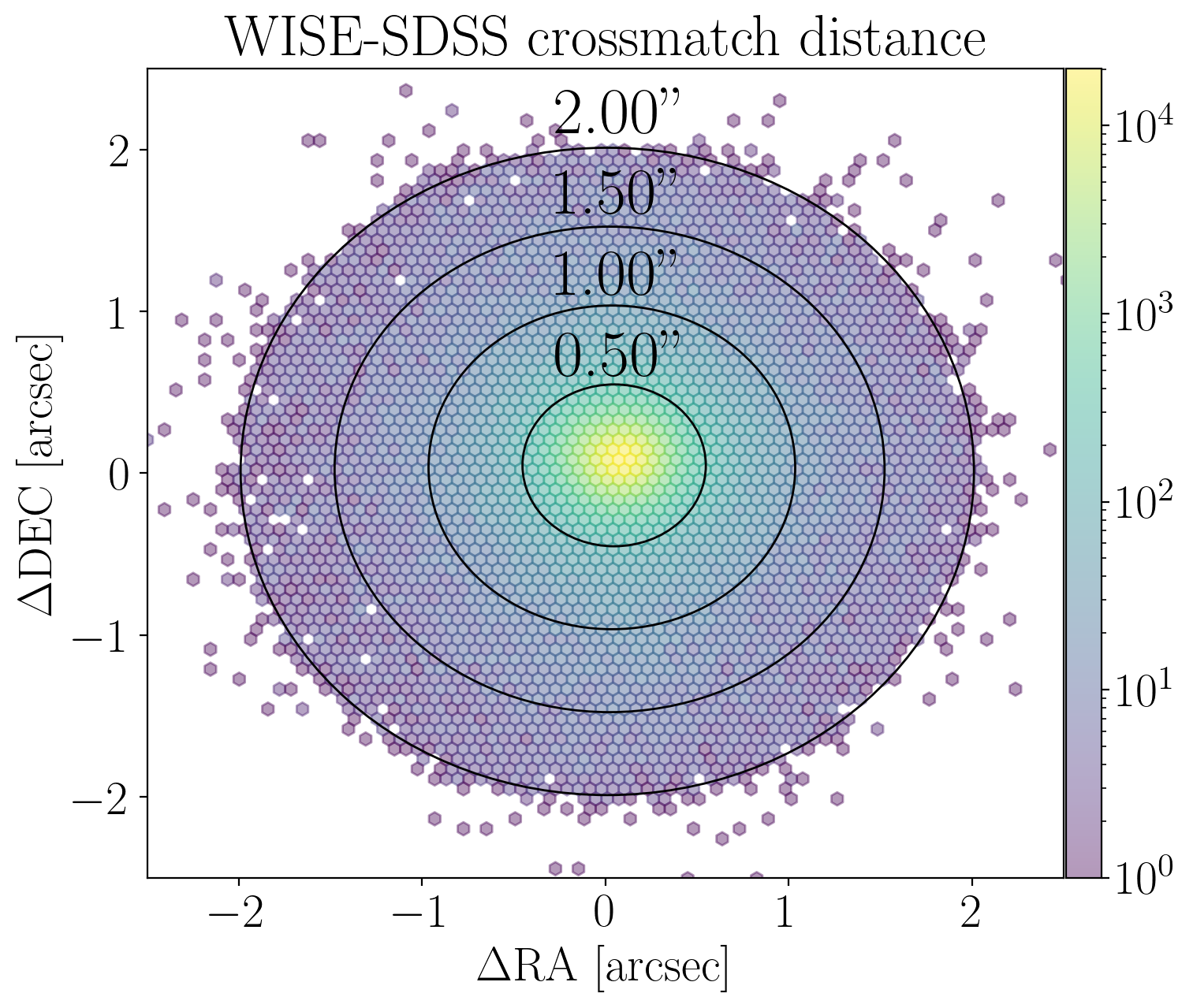}
  \caption{The astrometric distance match between WISE and SDSS sources in logarithmically colored bins. Almost all ($>99\%$) of sources are matched to within half an arcsecond, an excellent crossmatching accuracy. Only a few sources are outside the 2 arcsecond limit. This is only done for very bright/extended sources with no other possible match nearby.}
  \label{fig:poserr_crossmatch}
\end{figure}

The WISE - SDSS crossmatching utilised here is taken from the \href{https://skyserver.sdss.org/dr16/en/help/browser/browser.aspx?cmd=description+WISE_allsky+U#&&history=description+WISE_allsky+UURL}{SDSS skyserver webpage}.\footnote{Downloaded on 2024-02-26.} The crossmatching uses the original WISE photometry, unlike the forced photometry at SDSS astrometric positions of \cite{Lang2016_wise_sdss}. Since the W1 and W2 bands are more sensitive than W3 and W4, the used sources are all required to be 5$\sigma$ detections in these two bands \citep{Wright2010_WISE_overview, Lin2013_wise_sdss}. This requirement is still in place regardless of possible 5$\sigma$ detections in W3/W4, meaning that a galaxy may be well-detected in W1/W3/W4, but if it is not detected in W2, it is not included. This niche scenario fortunately includes only a very small subset of galaxies. Since W1/W2 are especially good at detecting massive, red, early-type galaxies out to large redshifts, the W1/W2 requirement preferentially select this type of galaxies, although the vast majority of star-forming galaxies are still included. 
Just like the detection cut in W1/W2 is quite harsh, so is the astrometric cut of 2 arcseconds. The distribution of astrometric crossmatch differences are shown in Figure \ref{fig:poserr_crossmatch}, where it is clear that almost all matches are better than half an arcsecond. A few galaxies with strong detections are included beyond the 2'' boundary if there is no other viable source nearby.

To crossmatch XMM-Newton with SDSS and WISE, we utilise the common software TOPCAT \citep{Taylor2005_TOPCAT} to select all sources within a common radius of 10'', which we then later restrict further to 5''. We furthermore require there to be no extended sources, since extended X-ray sources are unlikely to be AGN, but instead hot intra-cluster gas. The used XMM-Newton catalogue is the XMM-Newton serendipitous survey 4XMM-DR14 catalogue ({\href{https://heasarc.gsfc.nasa.gov/W3Browse/xmm-newton/xmmssc.html}{https://heasarc.gsfc.nasa.gov/W3Browse/xmm-newton/xmmssc.html}}).\footnote{Accessed on 2024-05-13}

\subsection{Training, validation and testing sets}
\label{appsec:train_test_split}

The fundamental ethos/validation of data-driven models is the ability of the model to describe data that was not used for constructing the model \citep{kuhn2013_applied, Hogg2024}. We therefore split our data into three sets, a training set, which is used for fitting the data driven models, a validation set, used for evaluating model hyperparameter decisions, and a test set, reserved for one final evaluation of the model performance before publication. The test is never used before this final and immediately pre-publication step. We use a 70/10/20 percent split for these sets.

\section{\spender}
\label{appsec:autoencoder}

\cite{Melchior2023_spender} introduced \spender, an autoencoder designed to work on galaxy spectra. \spender~comprises a convolutional encoder that compresses spectra into a low-dimensional latent space. This is followed by a decoder that reconstructs a rest-frame representation which may even include a broader spectral range and higher resolution than the original observation. The model applies explicit transformations, including redshift adjustment, resampling, and convolutions, to align the reconstructed spectra with the observed data. \spender~can process galaxy spectra at various redshifts and is resilient to artifacts such as residuals from skyline subtraction \citep{Jespersen2024_ragnar, Jespersen2024_SuNSS_SPIE}, allowing it to handle large survey data without extensive preprocessing. \spender~is trained on the entire spectroscopic galaxy sample from SDSS-II. \cite{Liang2023_spender} forced the latent space representation to be redshift-invariant by adding novel loss terms during training. These loss terms explicitly link distances in the latent space to those in the data space, preserving locality in the latent space. This adjustment resulted in a redshift-invariant, non-degenerate latent space distribution, effectively distinguishing between common and anomalous data. The improved model was able to identify outliers such as blends with foreground stars, highly reddened galaxies, galaxy pairs and triples, and stars misclassified as galaxies.

\section{Priors for \prospector}
\label{appsec:prospector}

\prospector~employs a comprehensive model of galaxy SEDs. The priors used for \prospector~can be found in Table \ref{tab:prospector_params_priors}. \prospector~utilizes a non-parametric SFH model, dividing the galaxy's history into distinct time bins. The non-parametric SFH model is detailed in \cite{LejaCarnall2019_measuring_SFHs_NONparametric}. The emission from the galaxy is shielded by dust. The dust absorption is parametrized by  three parameters. The two first are the diffuse dust optical depth, which quantifies the attenuation of light due to diffuse interstellar dust affecting \textit{all} stellar populations, and the birth-cloud to diffuse dust optical depth ratio. This parameterization is based on the two-component dust model described in \cite{CharlotFall2000_dustlaw}, which separates the attenuation effects of birth clouds and the diffuse interstellar medium. The third is the power-law slope of the dust attenuation curve. There is also significant dust \textit{emission}, described by three main parameters. The first, $U_{\min}$, specifies the minimum intensity of the interstellar radiation field that heats the dust, influencing the infrared emission. The second, $\gamma_{\mathrm{dust}}$, controls the fraction of dust heated by high-energy (starlight) radiation, affecting the warm dust emission component. The third is the PAH mass fraction, which is given in percent. All of these are modelled as part of the model of \cite{Draine_Li_2007_PAH}.
The nebular emission strength is largely governed by the radiation field strength for gas excitation $U_{\mathrm{gas}}$. The AGN component originates from \cite{Nenkova2008_AGNtorusI}, and is governed by $\mathrm{f_{AGN}}$, the ratio of the active galactic nucleus (AGN) bolometric luminosity to the total (AGN plus stellar) luminosity, and the optical depth of the dust torus surrounding the AGN, $\tau_{\mathrm{AGN}}$.

\begin{table*}[t!]
  \centering
  \caption{A list of parameters and priors used for \prospector. The selections is based on the \prospector-$\alpha$ model.}
  \label{tab:prospector_params_priors}
  \begin{tabular}{l|c|c}
  \hline
  \hline
   Parameter & Parameter Description & Prior \\
   \hline
  SFH & Star formation histories in bins of $\{[0-100 \mathrm{Myr}], [100-300 \mathrm{Myr}], $  & Beta(min = 0, max = 1, $\alpha$ = [5, 4, 3, 2, 1],\\
  & $[300 \mathrm{Myr}-1 \mathrm{Gyr}], [1 -3 \mathrm{Gyr}], [3 - 6 \mathrm{Gyr}], [6 - 13.6 \mathrm{Gyr}]\}$ & $\beta$ = [1, 1, 1, 1, 1])\\
  \hline
    $\mathrm{M_*}$ & Total stellar mass formed & LogUniform(min = $10^8$, max = $10^12$) \\ 
 \hline 
 $\mathrm{log(Z)}$ & Stellar metallicity & Uniform(min = -2, max = 0.19) \\ 
 \hline
  $\mathrm{\tau_{\lambda,1}}$ & Optical depth of diffuse dust affecting all light & Uniform(min = 0, max = 4) \\ 
 \hline 
   $\mathrm{r_{young}}$ & Ratio of optical depth of birth-cloud and diffuse dust  & ClippedNormal($\mu = 2$, $\sigma = 0.3$, min = 0, max = 2)\\ 
 \hline 
   $\mathrm{dust_{index}}$ & Power law index of the attenuation curve &  Uniform(min = -2, max = 0.5) \\ 
 \hline
  $\mathrm{U_{min,dust}}$ & Minimum radiation field for dust heating & Uniform(min = 0.1, max = 25) \\ 
 \hline 
  $\mathrm{\gamma_{dust}}$ & Relative contribution of dust emission from dust  &  LogUniform(min = 0.001, max = 0.15)\\ 
  & heated by high and low energy radiation & \\
 \hline 
  $\mathrm{q_{PAH}}$ & Percent mass fraction of dust in PAHs & Uniform(min = 0.5, max = 7) \\ 
 \hline 
   $\mathrm{U_{gas}}$ & Radiation field strength for gas excitation & Uniform(min = -4, max = 0) \\ 
 \hline 
   $\mathrm{f_{AGN}}$ & Ratio of bolometric luminosity of AGN to stellar component & LogUniform(min = $10^{-5}$, max = 3) \\ 
 \hline 
   $\mathrm{\tau_{AGN}}$ & Optical depth of AGN dust torus &  LogUniform(min = 5, max = 150) \\ 
  \hline
  \hline
  
  \end{tabular}
\end{table*}

\section{Science with latent spaces and overparametrization of models}
\label{appsec:science_with_latents_overparametrization}

Here we discuss some more tentative aspects of our work. We will discuss the implications, benefits and drawbacks of doing science starting from a latent space, the low dimensionality of this latent space, as well as the implication of overparametrization of galaxy models.

\subsection{Science with latent spaces}
\label{subsec:science_with_latents}

One interesting aspect of the approach taken in this paper is that the majority of the analysis is done starting from a low-dimensional encoded latent space, and not the SDSS spectra themselves. This discussion has specifically been avoided until now, since the main point of this work is not whether or not science can be done starting from latent spaces, but given that this topic has been a subject of discussion recently \citep{Hogg2024}, a small discussion is in order here. We have tested training an empirical model starting from the raw spectra, however, since the dimensionality of the inputs go from 8 to 4000, training becomes significantly more unstable, and we actually obtain a less accurate model than the one starting from the \spender~latent space. Another aspect worth discussing is the dimensionality of the \spender~latent space, since the fundamental dimensionality of the global galaxy hyperplane has been a popular topic ever since the discovery of global scaling relations like the fundamental plane for elliptical galaxies \citep{DjorgovskiDavis1987_fundamental_plane, Hyde2009_fundamental_plane} and the star-forming main sequence \citep{Speagle2014_SFMS}. The dimensionality of the \spender~latent space is six, which naively would suggest that the fundamental hyperplane has the same dimensionality. Although this enticingly suggests that not only cosmology, but also galaxy evolution, is nothing but a search for six numbers, one must be cautious when interpreting this dimensionality. This is due to the fact that arbitrarily complicated functions can theoretically be represented by single-parameter functions, \textit{if} the single parameter has infinite numerical precision \citep{Piantadosi2018_one_parameter}. This echoes the classic problem of ``Von Neumann's Elephant'', finding a four-dimensional curve which closely resembles the shape of an elephant, originating from von Neumann's comment ``\textit{With four parameters I can fit an elephant, and with five I can make him wiggle his trunk}''. However, for Von Neumann's Elephant the functions are understood to not be arbitrarily complicated, and with integer or low rational coefficients, a constraint that is not placed on the \spender~latent spaces. The dimensionality is greatly reduced if one uses only photometry \citep{Cooray2023_photometric_reduction}, a fact that is reflected in an increase in $\chi^2_N$ from close to 1 to around 10 when predicting WISE photometry based only on ugriz photometry. This is discussed further in Appendix \ref{appsec:optical_photometry}. The fact that we can use an extremely simple neural network, starting from a mere six-dimensional latent space, also suggests that modelling any galaxy SED from the UV to the MIR, is inherently a very low-dimensional task. The low dimensionality most likely owes to a tight coupling of different galaxy components, but this also implies that most galaxy models are vastly overparametrized. The historical decoupling of parameters that should be coupled separability may have made its way into our current models, it has here been demonstrated that this makes them imprecise and biased. By having inference models explore unnecessarily large parameter spaces, they more computationally expensive than necessary, which should be of concern to models like \prospector~with spend $\approx10$ hours to fit a single galaxy.

As detailed in \S \ref{sec:conclusion}, linking latent spaces from different data modalities also represents a very interesting idea. This has already been explored in a supervised setting, but the more interesting question is linking a latent space from theory with one from data in an unsupervised (or weakly supervised) manner. Unsupervised manifold learning on these matched latent spaces could then help identify different galaxy groups \citep{Jespersen2020_tSNE_GRB, Portillo2020_VAE_spectrum, Dubois2022_unsupervised_spectra, Dubois2024_unsupervised_spectra}. 

\section{Predicting WISE using other input data}
\label{appsec:predictions_other_inputs}
In the sections below we describe the results obtained from predicting WISE photometry starting from different inputs than the full optical spectrum. We specifically test optical emission lines, optical spectrum, and $\texttt{prospector}$-derived properties, fitting in the optical as well as the optical and IR. 

\subsection{Predicting WISE using absorption/emission lines}
\label{appsec:lines_only_precision}

This multiscale occlusion model evaluation is done with a model which expects all lines, meaning that this does not uncover the full capability of what the model can do given either a single or few lines. This requires re-training the model. Since retraining the model is inexpensive, we test a model which uses only H$\alpha$ and H$\beta$, and one which also uses other metal lines. The full set of lines are those shown in Figure \ref{fig:feature_importance} (Balmer lines, [OI], [OII], [OIII], [SII], [NII]), and line intensities are measured in both emission and absorption with a Gaussian line profile. 

\begin{figure}
  \centering
  \includegraphics[width=0.45\linewidth]{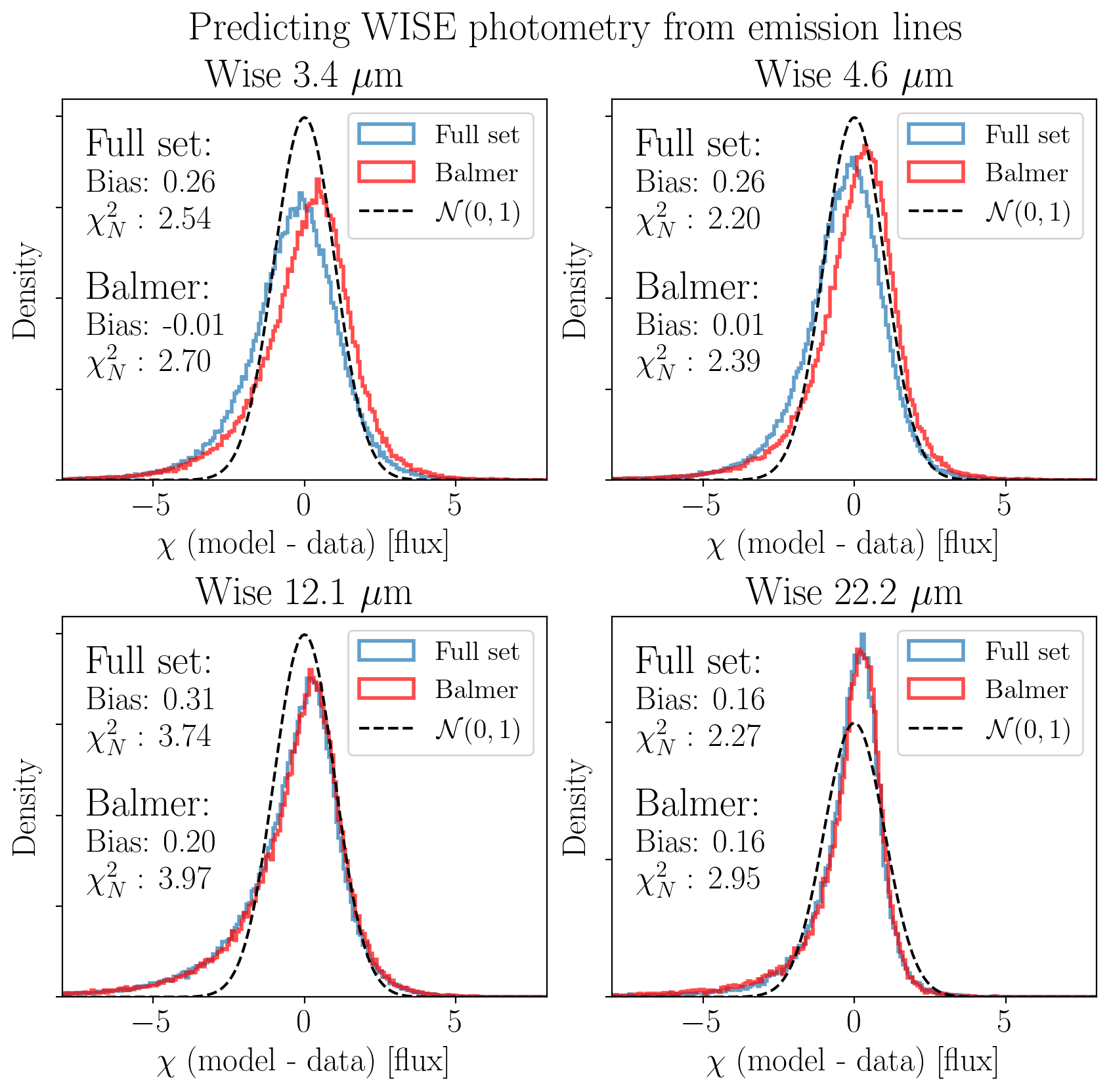}
  \includegraphics[width=0.45\textwidth]{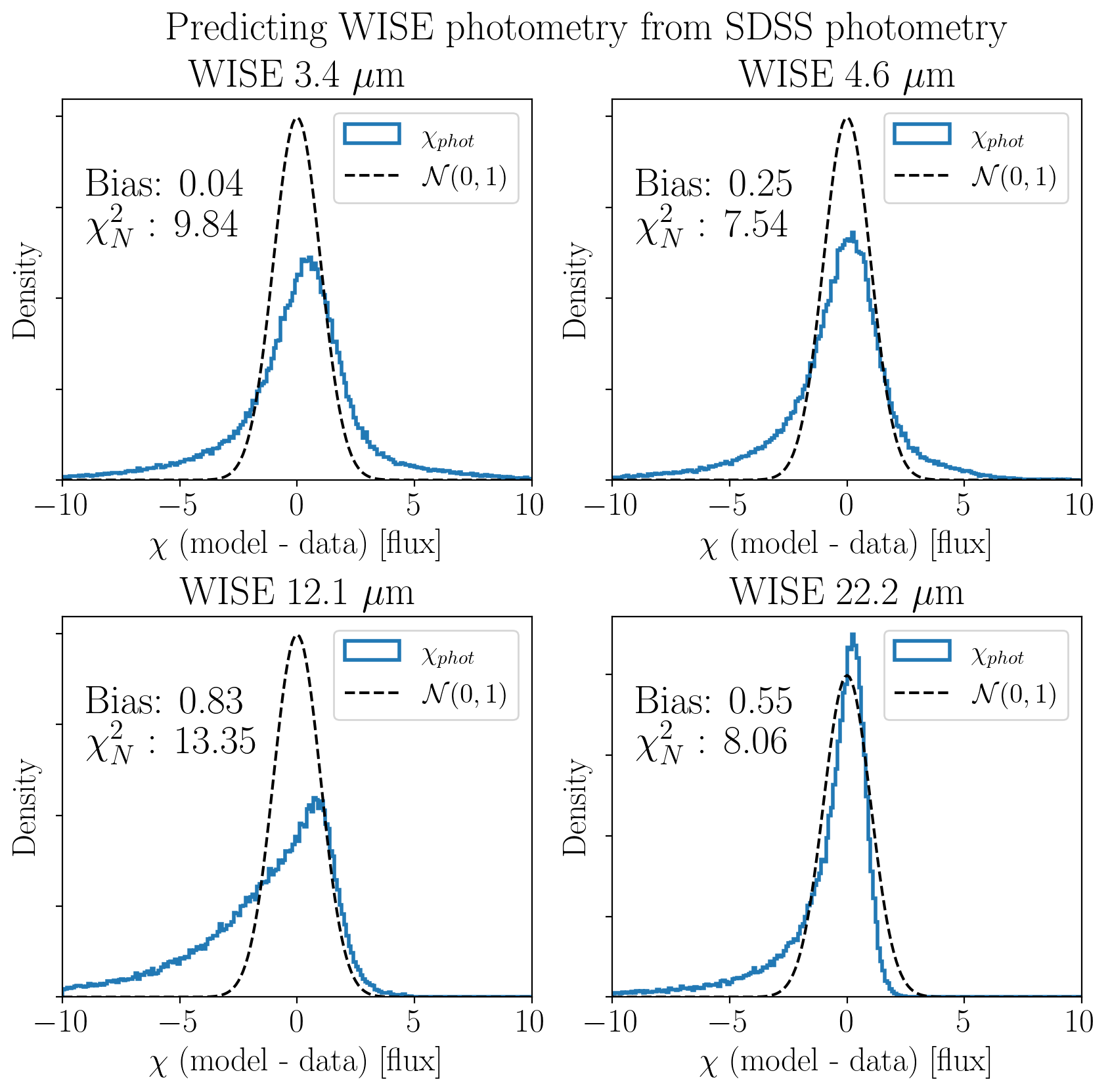}
  \caption{(left) Prediction made by an empirical model using only line intensities measured for the set of ionization lines shown in Figure \ref{fig:feature_importance}. We show results from using the first two Balmer lines and from using the full set of ionization lines. The performance of an empirical model using just Balmer lines is better than the tested SED models. However, the performance is still significantly worse than the full spectrum results. (right) $\chi$-distributions of the predictions of an empirical model trained to predict IR photometry given optical photometry and redshift, similar to Figure \ref{fig:chi2_N_detections}. The performance is an order of magnitude worse when using only the photometry, compared to using the full spectrum. The performance is still significantly better than that of SED fitting codes, but a similarity emerges, especially in the width of the distributions.}
  \label{fig:results_true_pred_balmer_ugriz}
\end{figure}

The results are shown in Figure \ref{fig:results_true_pred_balmer_ugriz}, which shows that the Balmer lines are indeed very important, but clearly do not uniquely drive the results shown in Figure \ref{fig:chi2_N_detections}, since the reduced $\chi^2$ is now twice as high. Adding oxygen, nitrogen, and sulfur lines improves the $\chi^2_N$, but does not bring it close to the results achieved using the full spectrum. We are clearly missing the full picture, presumably because the other ionization lines are not statistically independent and because we are missing important absorption lines as shown in Figure \ref{fig:feature_importance_zooms}.

\subsection{Predicting WISE using only optical photometry}
\label{appsec:optical_photometry}

The results presented in the main body of this paper suggest that the connection between the optical and infrared emission of a galaxy is tight, given that one has access to an optical \textbf{spectrum}. The strong connection between the optical and IR may not be inferable when one does not have access to a spectrum. To test this, we here train another MLP to map from ugriz SDSS photometry to WISE photometry, using the same methods as described in the main body of the paper. We use the spectroscopic redshift of the source.
The results are shown in the right panel of Figure \ref{fig:results_true_pred_balmer_ugriz}. It is clear that the optical-IR connection is now much harder to constrain, with $\chi^2_N \approx 10$ instead of $\chi^2_N \approx 1$. We can therefore conclude that optical photometry is insufficient to predict IR properties of galaxies. This, in addition to other reasons discussed in \S \ref{sec:discussion}, could also be one of the reasons for the poor performance of SED-modelling codes, since these codes are almost universally developed with a focus on photometry. It is important to stress that the difference in coupling strength between the photometry and the spectra is not something which is driven by our use of the \spender~latents. Autoencoding can never introduce additional information.

\subsection{Predicting WISE using $\texttt{prospector}$-derived properties}
\label{appsec:predict_from_prospector}

A natural question raised by our results is whether WISE photometry can be predicted to comparable precision using a reduced set physical parameters inferred from SED fitting, rather than directly from the optical spectra. To test this, we adopt as a baseline a simple empirical approach in which WISE magnitudes are predicted using a $k$-nearest-neighbor (kNN) regression, taking as inputs galaxy properties derived from \texttt{prospector} fits.

Specifically, we use the six star formation rate estimates (in different time bins), stellar mass, dust attenuation, and spectroscopic redshift inferred from $\texttt{prospector}$ fits to either the optical spectra alone or to the combined optical spectra+IR photometry. As a control, we compare against a null model that assigns every galaxy the mean WISE magnitude of the full sample.

We find that the use of SED-derived physical parameters leads to a clear improvement over the mean-magnitude baseline, demonstrating that $\texttt{prospector}$ does capture physically meaningful information relevant for predicting infrared emission. However, the achieved accuracy remains significantly worse than that obtained by our empirical model trained directly on the optical spectra. Moreover, we find no statistically significant difference between predictions based on properties derived from optical-only fits and those derived from optical+IR fits.

This limitation is partly driven by the restricted size of the training set, which cannot be substantially increased due to the high computational cost of $\texttt{prospector}$ fitting. More fundamentally, however, these results suggest that while SED-fitting compresses galaxy spectra into a compact and interpretable parameter set, this compression inevitably discards information that is predictive of mid-infrared emission. In particular, the model misspecifications identified throughout this work imply that MIR-relevant information is not efficiently captured by the inferred physical parameters, even when infrared data are included in the fit.

More generally, the experiment presented here can be viewed as a simplified test of the extent to which an SED model that embeds a strongly coupled prior between physical parameters and observables could achieve higher predictive accuracy for mid-IR photometry. In principle, a physically motivated, self-consistent model that encodes joint constraints on star formation histories, dust geometry, and emission processes could leverage such a prior to make more accurate predictions of WISE fluxes from optical data alone. Indeed, recent methodological work has highlighted the potential for coupled priors to guard against catastrophic inference errors in high-dimensional astrophysical inference problems \citep[e.g.,][]{Alsing2024_popcosmos}, suggesting a clear direction for future SED model development. However, the results of our current analysis also underscore that existing SED models suffer from significant misspecifications. Before the full promise of coupled priors for predictive accuracy can be realized, these underlying model inadequacies will need to be mitigated through improved physical fidelity and more flexible modelling frameworks.

\begin{figure}
    \centering
    \includegraphics[width=0.7\linewidth]{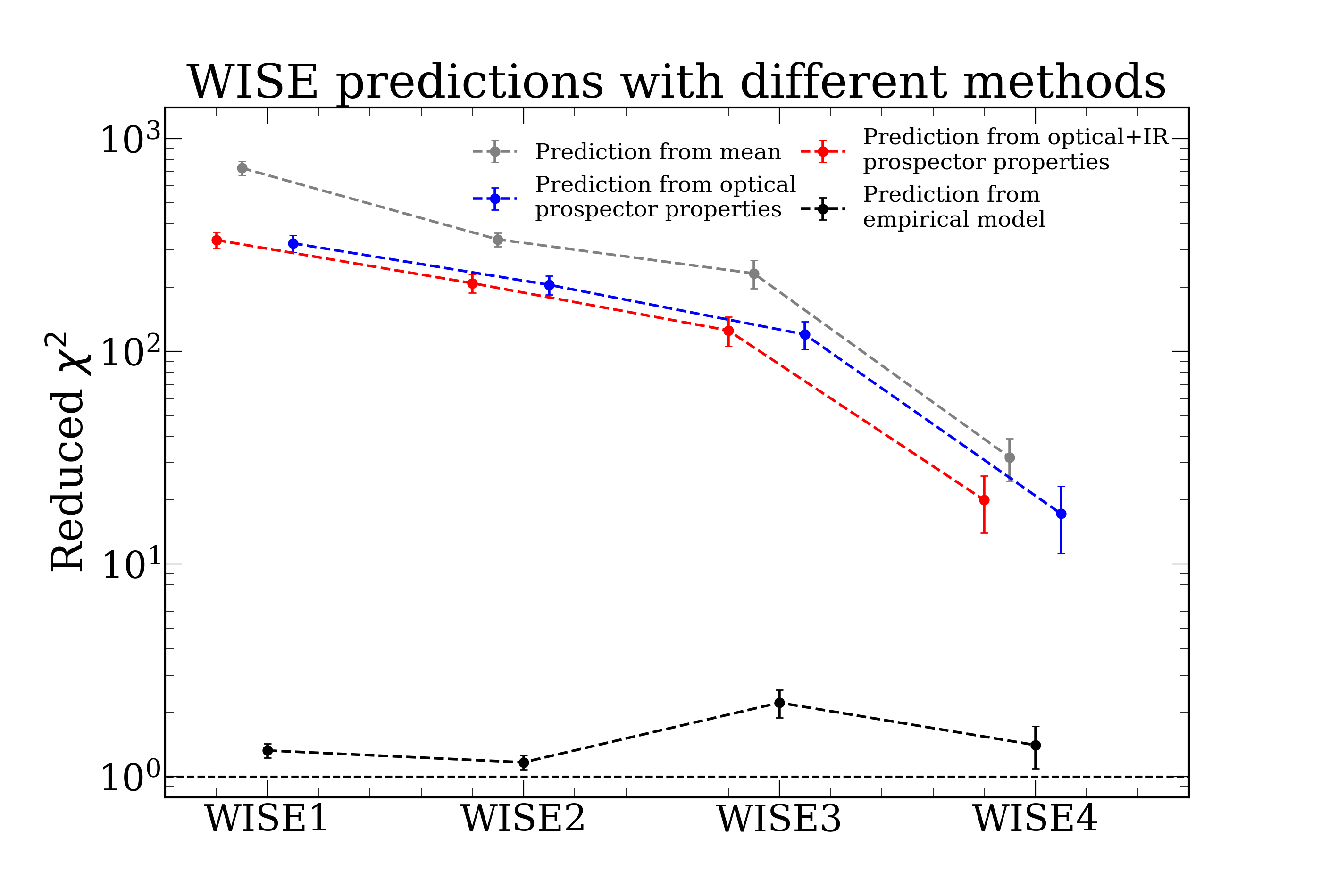}
    \caption{Prediction of WISE photometry using different inputs. As a baseline, we plot the prediction accuracy resulting from guessing that every galaxy has the mean magnitude of the full sample (gray). We plot the prediction accuracy of a k-nearest neighbor approach, taking as inputs the six SFRs, dust attenuation, and mass as derived from $\texttt{prospector}$ fits to the optical (blue) and optical+IR (red), as well as the spectroscopic redshift. After tuning k, we find that there is indeed a significant global improvement from having access to the derived physical properties, with no difference between optical only and optical+IR derived properties. However, these properties are not informative to the same extent of our empirical model (black). Uncertainties are computed by subsampling the galaxies used for training and testing.}    \label{fig:predict_from_prospector_properties}
\end{figure}

\section{Multiscale Occlusion}
\label{appsec:MO}

Our method, \textbf{multiscale occlusion}, is inspired by the work on \textbf{inverse multiscale occlusion} \cite{Shen2023_multiscale_occlusion} but with a different baseline. This change in baseline is necessary since \cite{Shen2023_multiscale_occlusion} investigated spectra already identified as outliers, aiming to identify sections of each spectrum making the spectrum abnormal. This implies that choosing a random other spectrum as a baseline will be a good choice, since any set of randomly sampled spectra will be more probably than a pre-identified outlier. 

\begin{figure}
    \centering
    \includegraphics[width=0.45\linewidth]{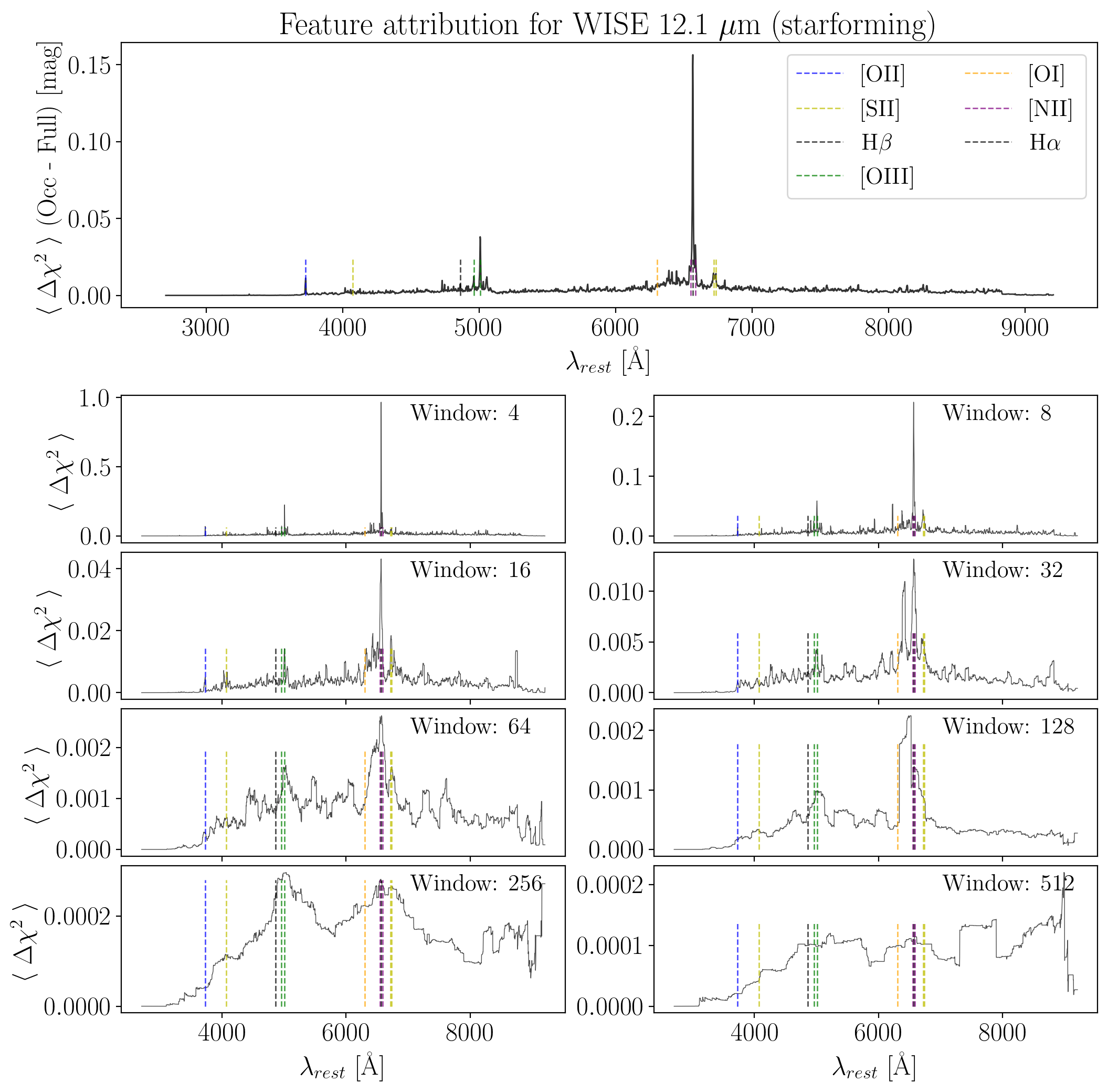}
    \includegraphics[width=0.45\linewidth]{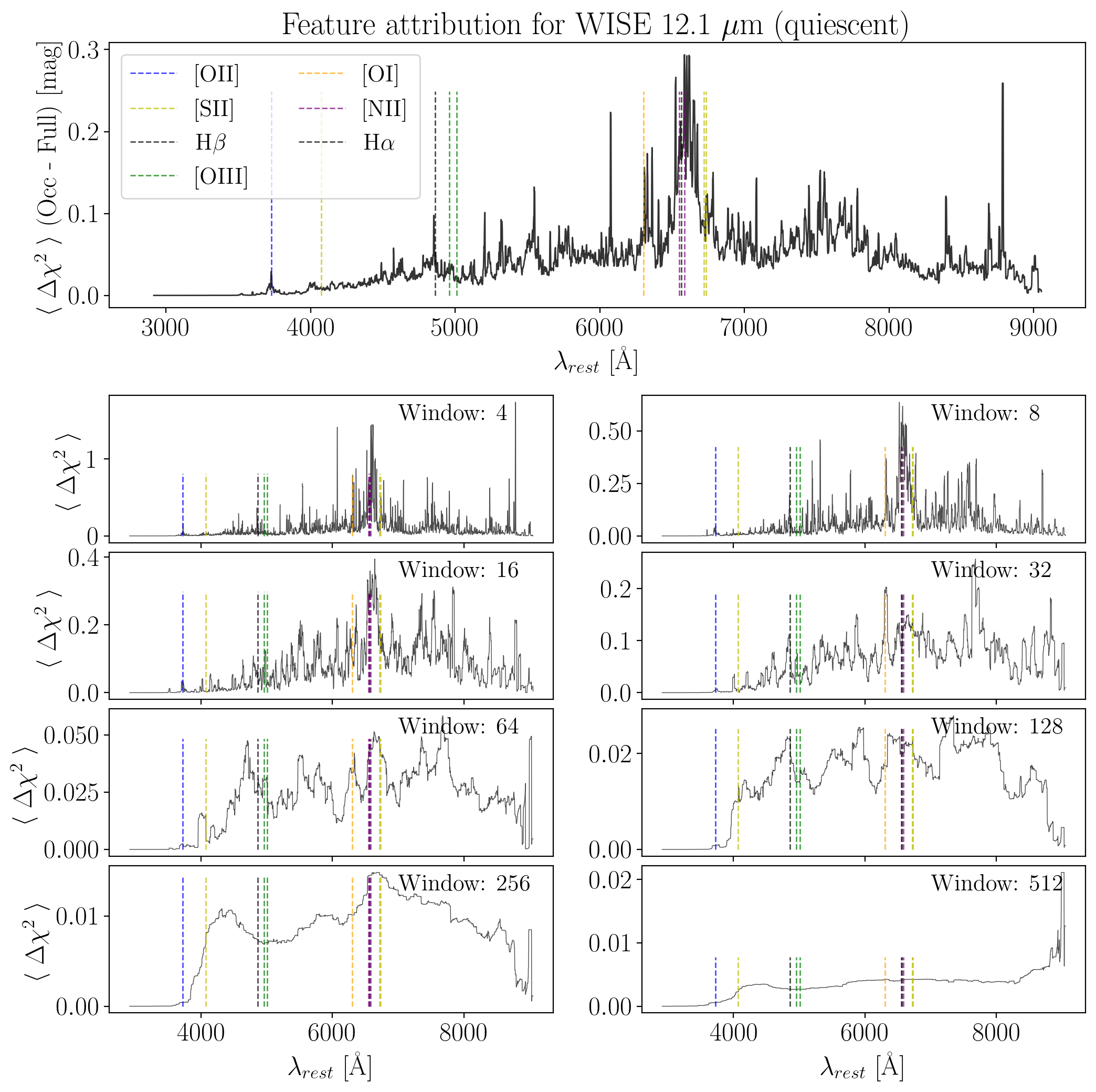}
    \caption{(left) Result of the multiscale occlusion for the W3 band for starforming galaxies. In contrast to Figure \ref{fig:feature_importance}, this figure shows the variance reduction (in $\chi$-space) due to a given pixel, not the bias. The vast majority of the variance in W3 can be explained by the intensity of H$\alpha$, [SII], [OII] and [OIII]. (right) Result of the multiscale occlusion for the W3 band for quiescent galaxies. In contrast to the situation for starforming galaxies (left), the information comes from all over the spectrum, not just a small set of ionization and metal lines. It is unclear what the features around 8700 Å are caused by.}
    \label{fig:feat_importance_W3_Q_SF_all_windows}
\end{figure}

However, here we are interested in analyzing the properties of the main galaxy sample, so choosing any other spectrum no longer makes for a valid comparison. Instead, we opt for replacing a spectral section of variable length N with noise, similar to if that section had been hidden by strong sky-lines. This way, our implementation of multiscale occlusion observes the response of our model to a spectrum with a given feature, compared to the model response of a spectrum without that feature. Choosing a range of lengths makes sure that we do not miss any important feature scales, however, it is somewhat of an \textit{ad hoc} approach, since the exact way to combine information from any given scale is unclear \textit{a priori}. Please see Algorithm 1 of \cite{Shen2023_multiscale_occlusion} for a pseudocode implementation of the inverse version of the multiscale occlussion algorithm implemented here.

\section{Splitting our sample -  brightness, mass, SFR, redshift}
\label{appsec:sample_split}

\begin{figure}
    \centering
    \includegraphics[trim={0.25cm 6.5cm 0.2cm 0.cm},clip,width=0.9\linewidth]{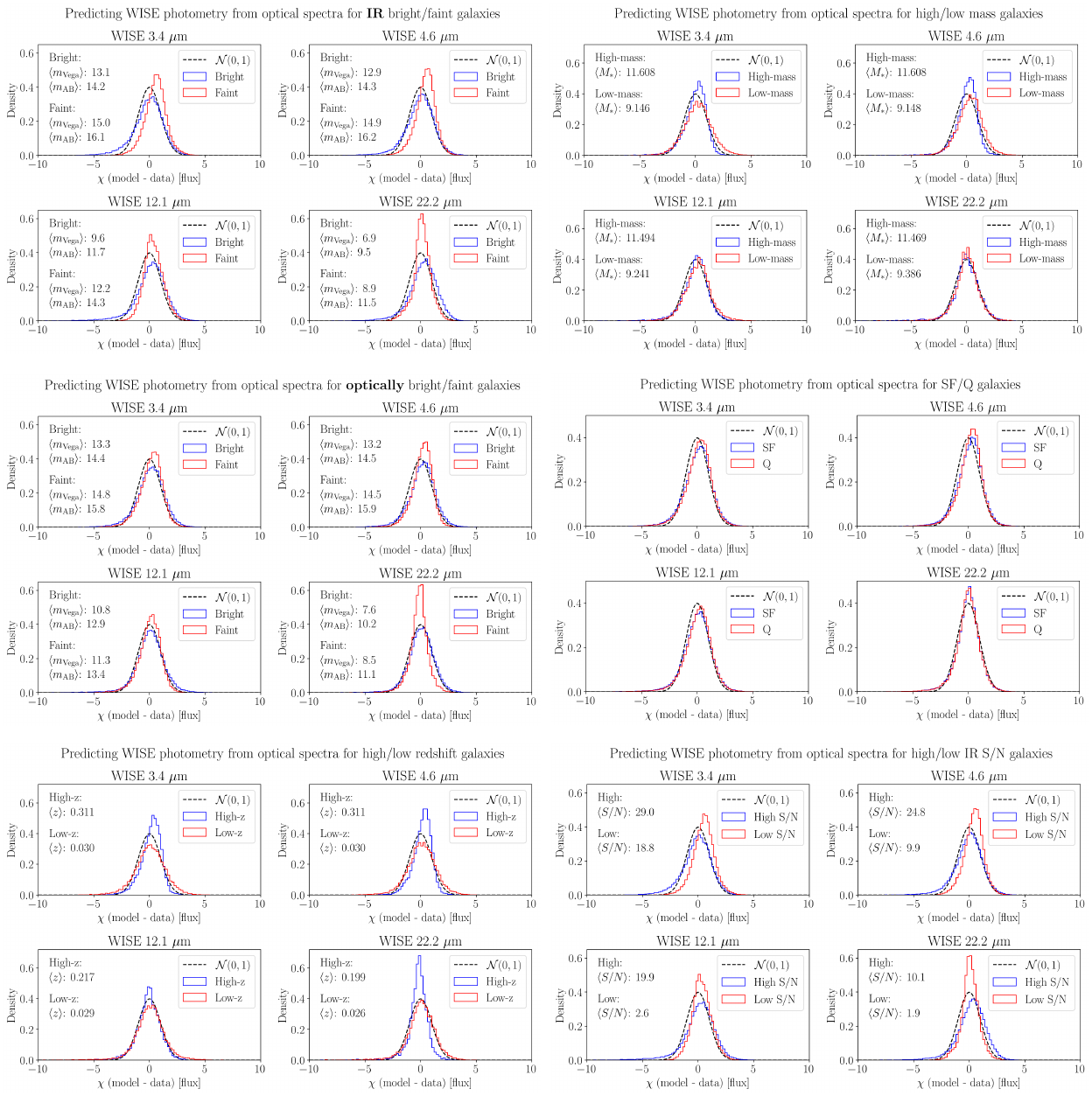}
    \caption{$\chi$-distributions of our predictions split by brightness in the IR/optical, mass, redshift, SFR and S/N. For all options, we show the highest/lowest 10 percent of the quantity in question. See the text for an expanded description of each quantity.}
    \label{fig:split_results_appendix}
\end{figure}

To make sure that our results are robust to changes in the underlying sample, we split our sample along 6 additional parameters on top of the ones done in the main text. These are shown in Figure \ref{fig:split_results_appendix}. Each split is done for the highest/lowest 10\% of the sample. We split along:

\begin{itemize}
    \item (top left) The brightness in each WISE band.
    \item (top right) The mass, as estimated using the methods of \cite{Bilicki2014_empiricalWISE} and \cite{Cluver2014_empiricalWISE}.
    \item (middle left) The average brightness in the optical.
    \item (middle right) Whether the galaxy is star-forming or quiescent.
    \item (bottom left) The redshift.
    \item (bottom right) The $S/N$ of each WISE band.
\end{itemize}

\section{Optical fit quality by wavelength}
\label{appsec:optical_fit_quality}

\begin{figure}
    \centering
    \includegraphics[trim={0.0cm 0.0cm 0.0cm 0.0cm},clip,width=0.95\linewidth]{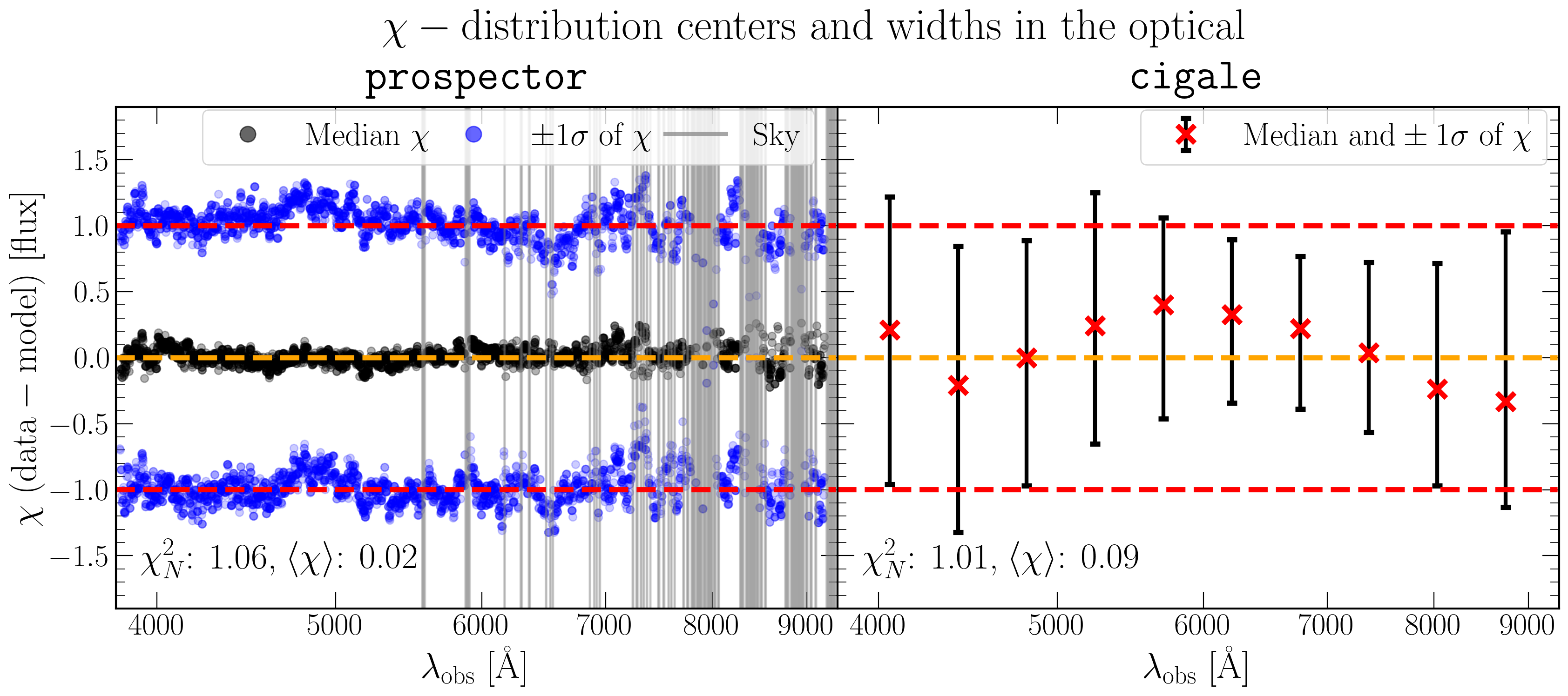}
    \caption{Width and center of the $\chi$-distributions fits in the optical. If perfect, the distribution would be centered at 0 (orange line) and have $\pm1\sigma$ widths of $\pm1$ (red lines). For perfect fits, the $\chi^2_N$ should also be close to 1, and the average $\chi$ should be close to 0. The fits are very good in general.}
    \label{fig:optical_fit_quality}
\end{figure}

Figure \ref{fig:optical_fit_quality} illustrates the quality of our spectral fits in the optical range by presenting the widths and centers of the $\chi$-distributions. Ideally, a perfect fit would result in a distribution centered at 0 with $\pm1\sigma$ widths of $\pm1$, along with global reduced chi-square close to one, and the average $\chi$ value close to 0. Our results demonstrate that the fits are generally very good, as the $\chi$-distributions align almost perfectly with these ideal expectations. However, despite the $\chi$-distribution metrics, the model does not extrapolate well beyond the fitted wavelength range. This discrepancy suggests that the issue lies not in the model's flexibility or its ability to fit the data within the observed range, but rather in its capacity to model the physics of galaxies accurately. The current model does not capture all the necessary physical processes or complexities required for accurate extrapolation. As demonstrated by \cite{Tachella2023_fitting} (Figure 4), the issue actually goes even further, with \prospector~not being able to predict the strengths of masked emission lines within the fitted wavelength range, \textit{even if all other lines from the same system (e.g., Balmer) are well fitted}. These models can thus fit data well, but the lack of extrapolation ability suggest that the results are highly dubious.

\section{The curious case of Arp220}
\label{appsec:arp220}

\begin{figure}
    \centering
    \includegraphics[trim={0.0cm 0.0cm 0.0cm 0.0cm},clip,width=0.8\linewidth]{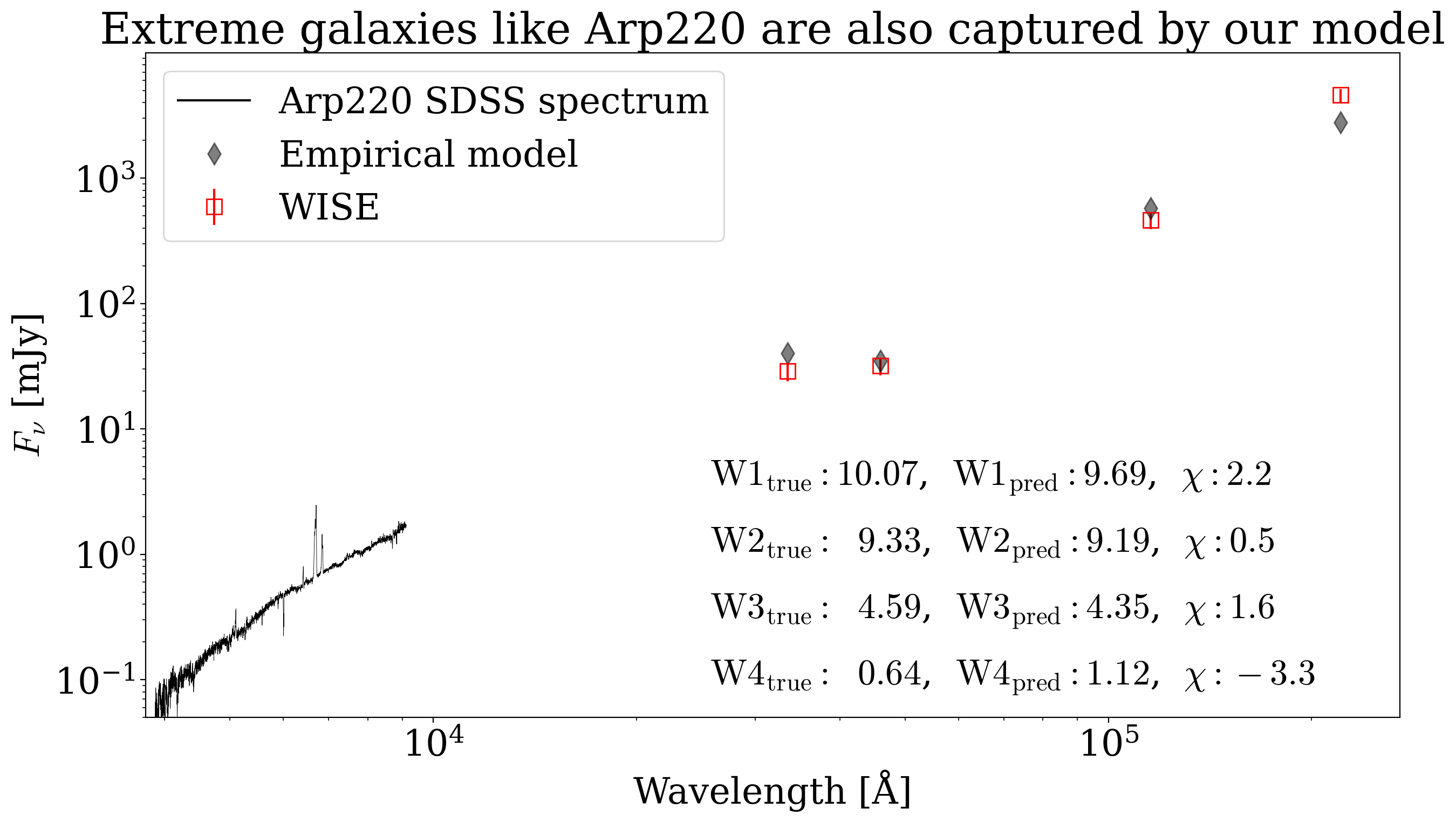}
    \caption{SDSS spectrum of Arp220, a nearby ULIRG, along with WISE photometry and the predictions of our empirical model. Typically, the deviations are on the order of 1$\sigma$, with only W4 being slightly underpredicted.}
    \label{fig:arp220}
\end{figure}

An obvious question is to what extent extreme outlier galaxies, e.g. the Markarian or Arp galaxies, are captured by our model. We therefore test our model specifically on Arp220, the closest ULIRG, which is not part of the training set. The results are shown in Figure \ref{fig:arp220}, where we see that the predictions are consistent with the rest of the sample, although W4 is slightly underpredicted. Even extreme ULIRGs are thus captured by our model. This test demonstrates that extreme attenuation alone does not cause failure.

\end{document}